# Examining Ownership Models in Software Teams

## A Systematic Literature Review and a Replication Study

**Umme Ayman Koana · Quang Hy Le · Shaikur Raman · Chris Carlson · Francis Chew · Maleknaz Nayebi**

**Abstract** Effective ownership of software artifacts, particularly code, is crucial for accountability, knowledge sharing, and code quality enhancement. Researchers have proposed models linking ownership of software artifacts with developer performance and code quality. Our study aims to systematically examine various ownership models and provide a structured literature overview. Conducting a systematic literature review, we identified 79 relevant papers published between 2005 and 2022. We developed a taxonomy of ownership artifacts based on type, owners, and degree of ownership, along with compiling modeling variables and analytics types used in each study. Additionally, we assessed the replication status of each study. As a result, we identified nine distinct software artifacts whose ownership has been discussed in the literature, with "Code" being the most frequently analyzed artifact. We found that only three papers (3.79%) provided code and data, whereas nine papers (11.4%) provided only data.

Using our systematic literature review results, we replicated experiments on nine priority projects at Brightsquid. The company aimed to compare its code quality against ownership factors in other teams, so we conducted a replication study using their data. Unlike prior studies, we found no strong correlation between minor contributors and bug numbers. Surprisingly, we found no strong link between the total number of developers modifying a file and bug counts, contrasting previous findings. However, we observed a significant correlation between major contributors and bug counts, diverging from earlier research. This study provides a comprehensive overview of ownership models and lists the variables used for ownership modeling in software engineering. Additionally, the study's findings can inform the development of best practices for software development teams and assist in decision-making, considering various company and project contexts.

Umme Ayman Koana
York University
E-mail: koana@yorku.ca
Quang Hy Le
York University
E-mail: lequang3@my.yorku.ca
Shaikur Raman
York University
Chris Carlson
Brightsquid
E-mail: chris.carlson@shaw.ca
Francis Chew
Brightsquid
E-mail: francis@brightsquid.com
Maleknaz Nayebi
York University
E-mail: mnayebi@yorku.ca

**Keywords** Ownership · Authorship · Software quality · Systematic literature review · Software Engineering · Code Ownership · Replication

## 1 Introduction

Ownership is the act, state, or right of possessing something [94]. In the field of Software Engineering, the concept of ownership, specifically pertaining to software code artifacts, has been a subject of discussion in various contexts, often tied to diverse performance metrics across different levels, including the project [81], product [35], and team [90] levels. Identifying the responsible developer for developing and maintaining software code has been a focal point in research, especially as developers' productivity [3] [52] [54] and the analysis of code quality have garnered increasing attention in the field. Yet, there is no systematic review of the existing models and the status and availability of the replication packages, benchmarks, and best practices. The lack of a standard benchmark makes it challenging for researchers to validate and replicate previous findings, which are essential for advancing knowledge and ensuring the reliability of research. One key objective of this study is to offer a comprehensive overview of the software ownership landscape within the field of software engineering. Through a systematic literature review, we aim to provide future researchers with a consolidated overview of evidence-based practices and effective strategies pertaining to ownership.

The literature widely acknowledges the important role of identifying ownership in large-scale software projects for improving the quality of the software product [24], [83]. The terms "Ownership" and "Code Ownership" have been frequently referenced in various studies [7][21][83]. However, there is no one source that refers to and defines ownership. After an initial literature analysis, we found 28 different definitions of ownership pertaining to code, team, module, collective ownership, organizational ownership, module ownership, issue ownership, task ownership, build ownership, requirement ownership, code authorship, and bug authorship. Code authorship is closely associated with ownership as well. Code authorship is commonly understood as the process of identifying the author of a piece of software based on its source code [49][40]. Furthermore, with the introduction of issue-tracking and version-control systems, log analysis has simplified authorship identification in a repository. Version control systems, such as GitHub, log and track the complete history of changes to a codebase, including information about who made each commit, when it was made, and what changes were introduced [82]. Meanwhile, with the introduction of peer coding and review practices, and as personnel and code changes occur, a code author cannot necessarily be considered the sole owner of the code. Hence, code ownership is concerned with identifying the developer(s) responsible for maintaining a software component or code snippet over time.

In the context of this existing rich literature, and mainly in a collaborative project with a software company, Brightsquid[1], we were interested in evaluating the ownership status within the company and comparing it with the status in other organizations. Our aim was to assess the extent to which the process of assigning tasks, following agile principles, and voluntary task selection within a team impact code quality. Brightsquid is a provider of healthcare practice data security services. Brightsquid solutions are HIPPA[2] compliant, enabling secure messaging and large file transfer for dental and medical professionals. Their communication platform supports aggregating, generating, and sharing protected health information across communities of healthcare patients, practitioners, and organizations. Our partner company Brightsquid offers solutions that facilitate clinical communication among physicians, patients, and clinic support personnel. Their communication platform supports aggregating, generating, and sharing protected health information across communities of healthcare patients, practitioners, and organizations. The company's journey began in 2009 in Alberta, Canada.

Brightsquid has evolved to have a distributed development team. The company uses the agile Scrum methodology for software development, fostering collaboration among its international team members. The company's organizational structure is flat, with fluid teams structured around specific products. The development team is globally distributed, and a total of 52 developers have contributed to the projects. Brightsquid has been working on numerous projects to achieve its business goals. The company manages 39 separate software repositories at various life cycle stages and with different activity statuses. With the COVID-19 pandemic, the company experienced team dynamics and customer demand changes within the healthcare sector. Brightsquid has actively focused on improving the developer's experience within the team. The company recognizes ownership as a crucial aspect of developers' experience. It is interested in (i) comparing their status with other software teams and (ii) exploring techniques to balance developer collaboration and accountability within the team. This study investigates ownership status within the development team and its relationship with code quality. To facilitate a meaningful comparison for our partner company, we initiated a systematic review to identify the context and the extent to which each study is replicable. The diversity in the range of software artifacts and terminology and differences in the offered models motivated us to perform a systematic literature review. We planned to use the systematic literature review results to analyze and evaluate the status of Brightsquid. In particular, we are interested in answering the research questions (RQs) below.

[1] https://Brightsquid.com/
[2] HIPAA: Health Insurance Portability and Accountability Act of 1996

**RQ1 (Literature Review):** How is ownership defined and modeled in software engineering literature?

We gathered and synthesized definitions provided for ownership within state-of-the-art software engineering, along with the different attributes used to model ownership. For each paper, we collected the author's *definition* of ownership, the *level or granularity of ownership* (file, module, or line of code, etc.), and the type of *artifact* (code, documentation, commits, etc.) that ownership was studied. We then synthesized a comprehensive list of these attributes across the different papers.

**RQ2 (Literature Review):** What treatment and outcomes have been used in modeling ownership in software engineering literature?

While different studies consider a variety of artifacts and their ownership (**RQ1**), each study modeled ownership differently. We have compiled and synthesized a list of outcomes, such as the number of bugs and developers' productivity associated with each ownership model, including team and individual behaviors.

**RQ3 (Literature Review):** What analytics and research methods have been employed in the study of ownership in software engineering, and to what extent are these studies replicable?

To build upon existing research and establish a meaningful baseline for new methods, we seek to understand the extent and type of evaluations conducted on ownership models. In addressing **RQ3**, we summarize the *model type* (descriptive, diagnostic, predictive, or prescriptive), *nature of the project* (open source, industrial, etc.), *accessibility of data and tools*, and *reuse of the study* (whether the study replicated a previous model).

**RQ4 (Replication):** How does the code ownership status in Brightsquid compare with the findings of three selected state-of-the-art studies?

We intended to evaluate the status of code ownership in Brightsquid. To achieve this, we selected three studies as case study models and used methodologies similar to each for analyzing nine active projects within Brightsquid. In response to **RQ4**, we compare the reported results from the original studies with our observations from Brightsquid.

This paper describes our methodology for systematic literature review in Section 2 and benchmarking in Section 3. In Section 4, we present the results of our systematic literature review and answer **RQ1**, **RQ2**, and **RQ3**. In Section 5, we provide more details on the status of Brightsquid's team and code. We will report the results of our ownership replication analysis in response to **RQ4**. In Section 6, we discuss the implications of our work, and in Section 7, we list potential threats to the validity of our study. Finally, we conclude our study in Section 8.

## 2 Protocols Used for Systematic Literature Review

Among the guidelines for systematic literature reviews in software engineering [41], we adhered to the systematic literature review framework proposed by Kitchenham et al. [43], which is widely accepted in the field.An overview of the study selection process is provided in figure 1, and this process consisted of five main steps:

1. Defining research questions - to highlight a specific area of interest, the researcher defines the research questions. In this study, we have three research questions (**RQ1**, **RQ2**, and **RQ3**) as detailed in the previous section.
2. Searching for relevant papers - Researchers create a search string as a result of analyzing the research questions and use it to find relevant papers in scientific databases. For our systematic literature review, we collected papers from five databases commonly known and used by the research community: IEEE Xplore, Scopus, ACM Digital Library, ScienceDirect, and Inspec. In Table 1, we have listed the 17 search strings used to gather papers within the scope of this systematic mapping. To define these 17 search strings, we first identified keywords related to our research questions. Then, we used the Boolean operator "AND" to combine keywords and expand our search queries. From the selected databases, we collected a total of 44,605 papers using our search strings. Among these, there were 15,911 overlapping results between the searched databases.
3. Screening of papers by reading the titles and keywords - Two annotators reviewed the papers and discarded those that were irrelevant to the research area based on the paper titles and keywords. After applying the inclusion criteria, we were left with 25,501 papers. We then applied exclusion criteria, removing non-English papers and 15,911 papers redundantly found in multiple databases (as detailed in Table 2). This exclusion resulted in removing 15,911 papers, leaving us with 9,590 papers. To further narrow down our analysis, we performed title and keyword-based exclusion, eliminating papers that did not include any subjects relevant to ownership, authorship, productivity, quality, leadership, individual performance, team performance, team characteristics, type, or size in the title or keywords. This process yielded 277 papers.
4. Exclusion based on abstract - Two independent annotators reviewed the abstracts to assess whether the included papers aligned with the scope of the research questions. In cases where one of the researchers disagreed with inclusion, the authors held discussions to resolve the differences. If the two researchers could not reach a consensus, the last author acted as the moderator. To synthesize the papers, we developed a classification scheme based on our research questions to provide answers:
   - Is the paper related to the ownership of software artifacts?
   - Is the treatment and/or outcome of the paper related to a software product or project?
   - Is there any open-source repository associated with the paper or not?

   Two researchers independently read the full paper abstracts and categorized them as "Yes" or "No" for each of the above questions. We then excluded 130 papers that received a "No" answer for all three categories, resulting in 147 papers for full reading and review.

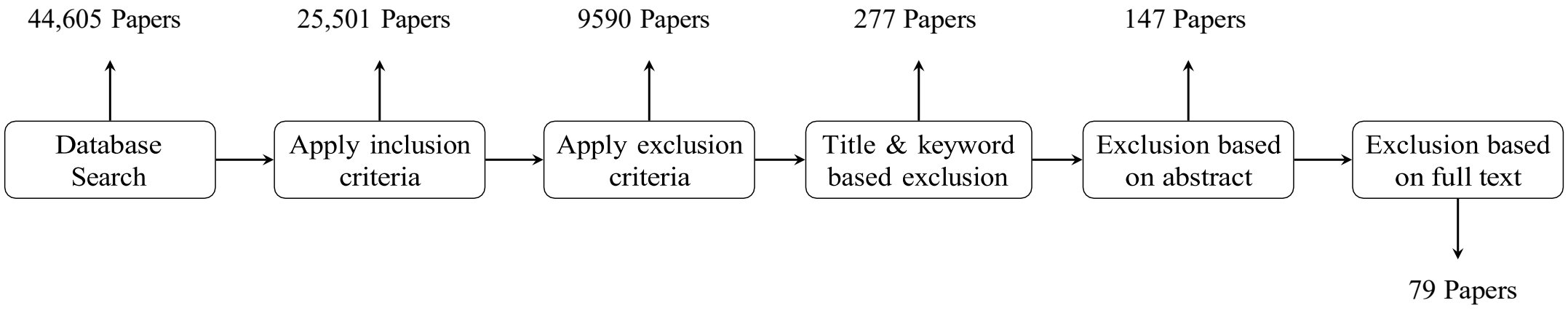

**Fig. 1** Study Selection process for our systematic literature review

**Table 1** Search string and number of returned records within each database

| ID | Search Term | IEEE | Scopus | ACM | Science Direct | Inspec |
|---|---|---|---|---|---|---|
| 1 | "Software Engineering" & "Ownership" | 482 | 389 | 3,407 | 2,383 | 767 |
| 2 | "Software Engineering" & "Ownership" & "Code" | 84 | 94 | 2,631 | 1,702 | 358 |
| 3 | "Software Engineering" & "Ownership" & "Developer" | 51 | 64 | 1,414 | 1,208 | 67 |
| 4 | "Software Engineering" & "Ownership transfer" | 24 | 3 | 82 | 32 | 19 |
| 5 | "Software Engineering" & "Authorship" | 176 | 92 | 720 | 3,900 | 245 |
| 6 | "Software Teams" & "Ownership" | 53 | 4 | 176 | 158 | 7 |
| 7 | "Software Teams" & "Ownership transfer" | 3 | 0 | 0 | 0 | 0 |
| 8 | "Software Development" & "Ownership" | 259 | 207 | 3080 | 4,042 | 353 |
| 9 | "Software Development" & "Authorship" | 52 | 34 | 581 | 2,956 | 50 |
| 10 | "Software Engineering" & "Requirements" & "Ownership" | 59 | 42 | 2,247 | 2,040 | 80 |
| 11 | "Software Engineering" & "User Requirements" & "Ownership" | 12 | 0 | 181 | 251 | 1 |
| 12 | "Software Engineering" & "Ownership" & "Productivity" | 19 | 18 | 791 | 760 | 38 |
| 13 | "Software Development" & "Ownership" & "Productivity" | 14 | 12 | 857 | 1,436 | 28 |
| 14 | "Software Engineering" & "Ownership" & "Code" & "Productivity" | 6 | 12 | 680 | 586 | 26 |
| 15 | "Software Engineering" & "Ownership" & "Software quality" | 80 | 23 | 479 | 287 | 45 |
| 16 | "Software Engineering" & "Ownership" & "Software quality" & "Productivity" | 7 | 3 | 215 | 152 | 5 |
| 17 | "Software Engineering" & "Software Development" & "Owner- | 28 | 11 | 410 | 241 | 14 |

During the abstract-based exclusion process, we excluded a set of papers that were relevant to the ownership of software's Intellectual Property (IP) among collaborating organizations, in open-source environments, and between different teams within the same organization [100, 33, 20]. We also excluded papers which were related to the *software design and object ownership* discussing the code or UML diagrams [42, 79, 15], *cloud based ownership* [95, 32, 40, 31], system level ownership [86, 85], *Non-Software document ownership*, for example, ownership of Wikipedia content [101, 56, 34]. These papers did not directly address ownership of software artifacts and received a "No" answer for at least two of the three questions above.

Exclusion based on full text- We read the full papers and made decisions regarding their inclusion. After reading the full text of those 147 papers, we identified only 79 papers that were directly related to our study. These 79 papers constitute the final set that we studied in depth.

5. Data extraction and mapping - In this final step, the first and the second authors independently reviewed the full papers, gathering and synthesizing information relevant to each research question. This step involves extracting all necessary detailed data for presenting and analyzing our research questions. During this stage, we recorded comprehensive data needed to answer the research questions, including ownership definitions, studied research questions, types of applied data analytics, dependent and independent variables used for modeling, empirical and evaluation setups, and replication status and availability.

The authors synthesized and organized the data to address the research questions by collecting all the information according to the described empirical protocol. We started by manually going through the papers and filling in the table provided in Appendix II to structure our findings. This table involves the dependent, independent, and control variables, the type of artifact being discussed, the subject, and the ownership model. Two authors independently filled out this table, followed by a discussion of any differences. The third researcher (last author) mediated any disagreement or ambiguity. As such, we built an inductive taxonomy [84] [80]. In Section 3, we present the results of our literature review in alignment with each research question.

**Table 2** Inclusion and Exclusion criteria applied in Step 2 if Systematic Literature Review in our study

| Inclusion Criteria | | Exclusion Criteria | |
|---|---|---|---|
| **IC1** | Studies related to Software engineering only | **EC1** | Studies not written in English |
| **IC2** | Studies published in journals/conference proceedings | **EC2** | Duplicated studies |
| **IC3** | Studies published from 2005 to 2022 | | |

## 3 Protocols Used for Benchmarking and Replication

Software engineering research has relied heavily on empirical studies and experimentation to advance its state of the art and practice. One crucial aspect of empirical research is replication, where an independent group of researchers externally reproduces the results of a study. Replication is widely regarded as a cornerstone of many disciplines, serving as an essential means of verifying and validating empirical studies [11] [23]. However, not all replication studies can be or need to be translated and discussed to verify the results of an initial study. Replications help researchers understand the impact and significance of contextual factors on the validity of empirical findings. Without proper replication, it is unclear whether study results are accidental, artifactual, or actually conform to reality [23].

In this context, we were interested in replicating the software development process within the partnered company. Our aim was not to verify the original studies but to leverage them as an established status quo and evaluate if their results were applicable in the context of Brightsquid. Thus, our goal was to provide a comparison to assess the quality of the development processes at Brightsquid. To replicate the studies in the partnered company, we followed the steps outlined below:

(i) Structuring the existing body of knowledge and identifying the setup of each experiment - We conducted a systematic literature review. We established a taxonomy for measuring and modeling ownership, focusing on aspects such as what, who, and how it has been assessed.

(ii) Identifying experimentation elements - We identified the dependent, independent, and controlled variables of each model following the general experimentation process as outlined by Wohlin et al. [99] as shown in Figure 2.

(iii) Evaluating if reuse is possible - To build on the existing knowledge and to reduce divergence between our study and the existing ones, we assessed if any replication material exists. This is mapped to **RQ3** of our study.

As a result of the systematic literature review, we could not identify reusable material (See Table 5) for analyzing the status of our partnered company. Therefore, in **RQ4**, we followed the following steps to replicate the selected studies.

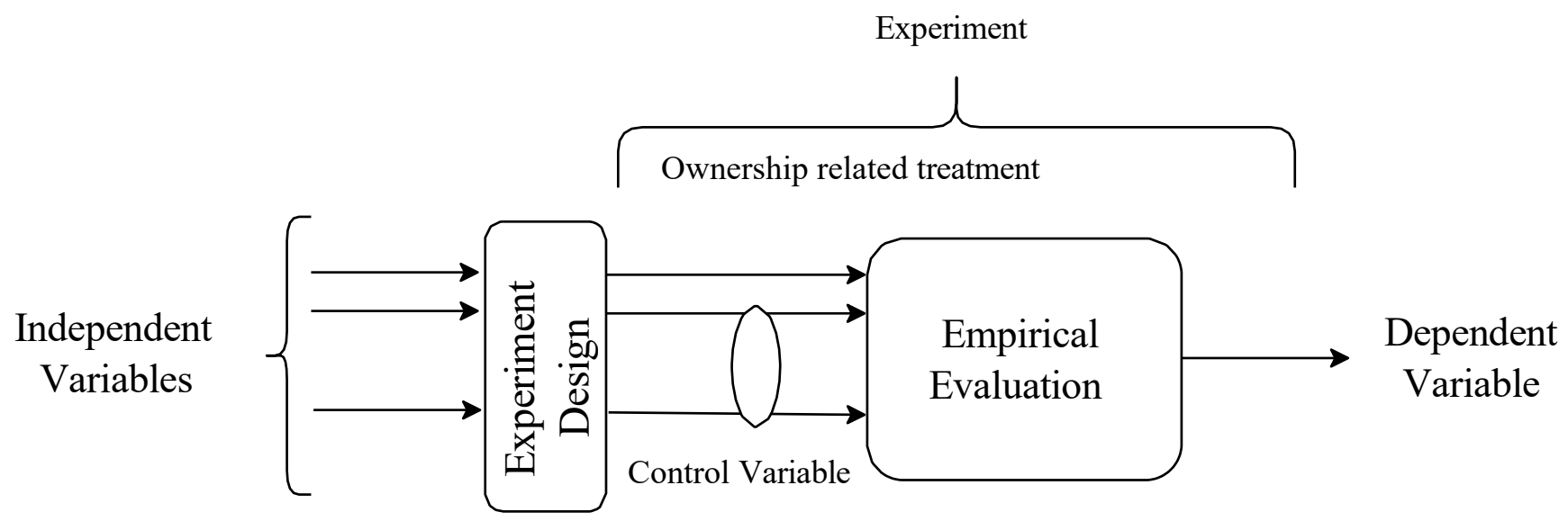

**Fig. 2** Treatment-outcome relation as we studied in this SLR. The process and the diagram are adopted from the work of Wohlin et al. [99].

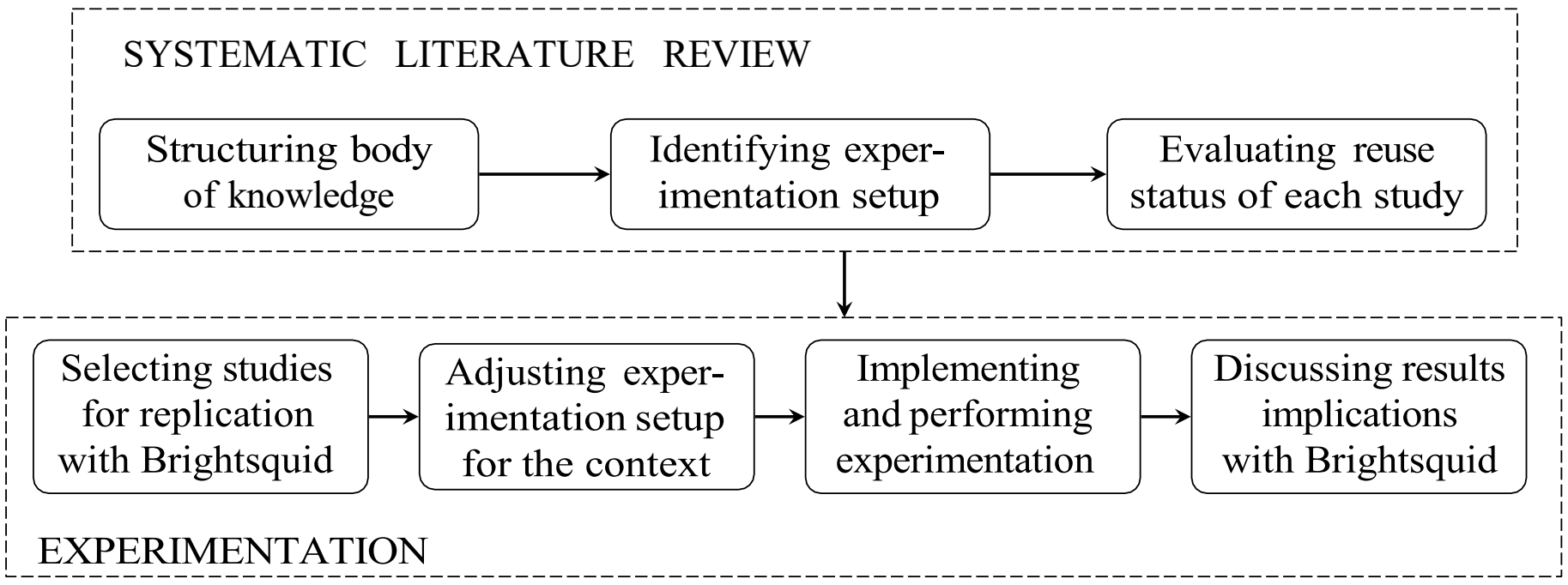

**Fig. 3** Process of replicating studies for Brightsquid

(iv) Selecting studies for replication at Brightsquid - After gaining an overview of all the studies, we discussed the context, meaningfulness, and importance of each ownership model within Brightsquid. Subsequently, we prioritized the studies for replication.
(v) Adjusting the experimental elements for our partner company data and code structure,
(vi) Implementing experimental elements and performing experimentation using Brightsquid data,
(ivv) Discussing the results in comparison with the original study and between the studies and identifying implications in collaboration with Brightsquid.

Our study methodology is summarized in Figure 3. Different replication types have been introduced for software engineering studies, including those discussed by Wohlin et al. [99] and Gomez et al. [23]. In our study, we performed a *close replication* by using the same research questions and closely following the experimental procedure. However, as with any non-exact replication, there were differences in factors such as the *site where the experiment was conducted*, *experimenters conducting the study*, *instrumentation and data*, and the *subjects involved* [99] [37]. Using the same taxonomy, we closely adhered to the experimental design and variables measured.

## 4 Result of Systematic Literature Review (RQ1, RQ2, RQ3)

Our literature review is based on a systematic search of 79 papers, which we thoroughly examined. These papers cover the period from 2005 to 2022. Figure 4-(a) presents the number of papers based on publication type, while Figure 4-(b) illustrates the yearly distribution of these 79 papers. Of the 79 papers, 45 were retrieved from conference proceedings, 28 were sourced from journals, and six were from workshops. We also observed that the highest number of papers (12) was published in 2021, with the second-highest number of publications in 2015.

To address **RQ2**, we evaluated every paper. We created a summary of the dependent, independent, and controlled variables used to model software ownership, as well as a summary of all definitions related to the topic. In the following sections, we provide a detailed discussion of our findings in response to each research question.

### 4.1 Ownership Definitions and Models (RQ1) - How is ownership defined and modeled in software engineering literature?

The definition and categorization of ownership are often determined by individual authors within the specific context of their studies, leading to inconsistency in the terminology used in this field. To improve clarity and comprehension, we adopted the most widely used terminology and provided cross-references across various studies. Upon reviewing these papers, we identified two main themes of discussion: Psychological ownership and Corporeal ownership.

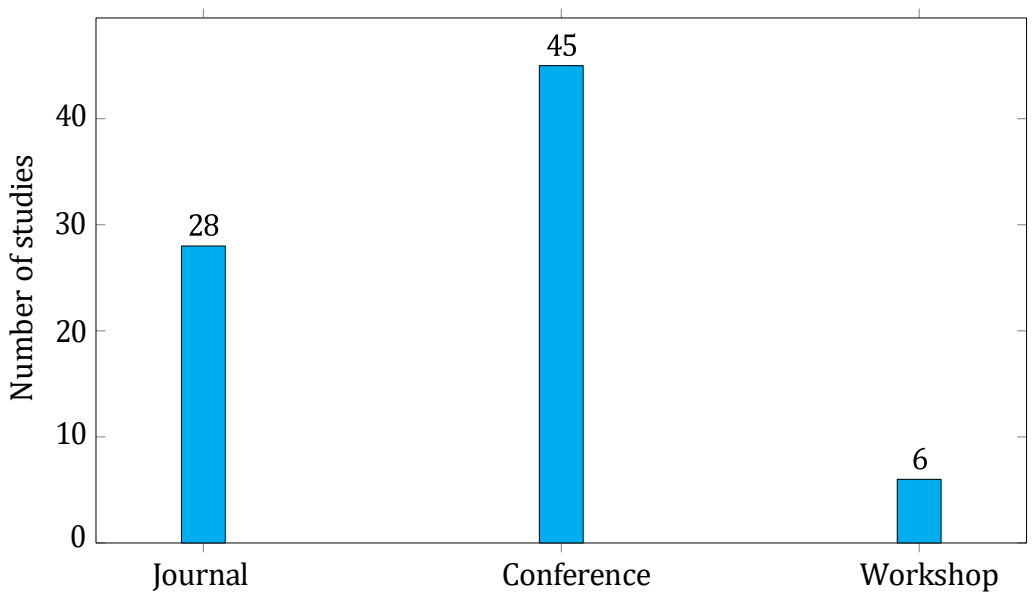

(a)

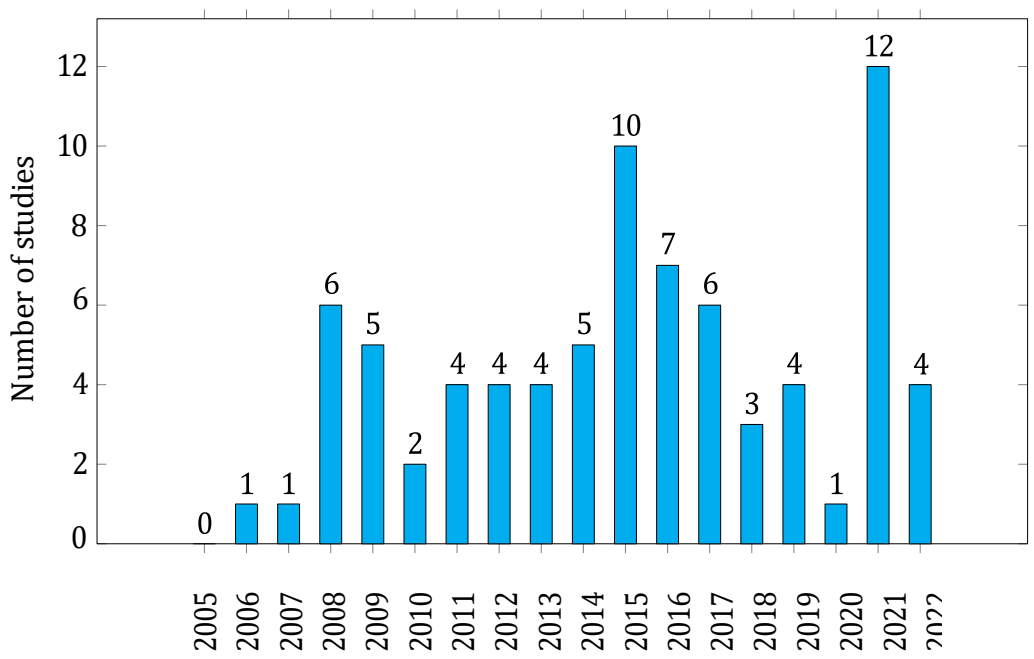

(b)

**Fig. 4** Distribution of paper based on (a) publication type and (b) publication per year.

**Psychological ownership** refers to the feeling of ownership that a person may develop toward an owned entity in the project [16].

**Corporeal ownership** refers to the development history of an artifact, often by referring to the Git history and contributions of the developers in the code base.

In the software engineering literature, the central focus revolves around different facets of what we term Corporeal ownership. According to Bird et al. [8], ownership refers to a general term used to determine whether a particular software component has a responsible developer or not. The ownership of a software artifact is subject to change based on the strategies and logistics a software team implements. It is commonly assumed in such studies that developers have the freedom to modify other teammates' solutions whenever necessary [96]. Consequently, the ownership of a software artifact is often shared among the developers. However, some studies focus on identifying an individual responsible owner for a particular artifact, while others discuss collective responsibility among team members. In this study, we present our findings about ownership from three different aspects to address **RQ1**.

**Type of artifact:** Refers to the artifact whose ownership is being discussed within a software team.

**Owner of the artifact:** Refers to the individual whose role in the maintenance of the artifact is being discussed.

**Degree of ownership:** Indicates the extent of rights and responsibilities of the owner, as well as the responsibilities of team members in maintaining an artifact.

In software engineering, some authors have used the terms authorship and ownership interchangeably [13] [53]. For example, Muller et al. [53] use the term code ownership to refer to the authorship of a particular software artifact; "The term code ownership is used in software engineering to describe who authored a certain piece of software." However, in our systematic review, we distinguish between authorship and ownership as two distinct yet related concepts. Authorship pertains to identifying the author of an artifact, such as a line of code. Bogomolov et al. [9] formulated the authorship problem as "given a piece of code and a predefined set of authors, to attribute this piece to one of these authors, or judge that it was written by someone else."

In contrast, ownership involves identifying the member(s) responsible for maintaining a software artifact, regardless of whether they originally authored it [97]. The collaborative nature of software maintenance and dynamic code changes often makes it challenging to attribute authorship to a single developer. Rahman and Devanbu clearly distinguished between the terms "authorship" and "ownership" [83]. They considered any developer who contributed to a code snippet as an author, designating the developer with the highest contribution as the owner of that code snippet. In our systematic literature review, we also differentiate between these two terms, acknowledging that ownership, at all levels, can be either shared or individual.

### *4.1.1 Type of Artifact (What?)*

Software artifacts are tangible products that result from development activities within software teams. Through our systematic literature review, we identified nine different artifacts whose ownership has been the subject of one or more studies. Table 3 provides a condensed definition of the ownership of each artifact as a result of our detailed synthesis in this systematic literature review.

Among all the artifacts, **"Code"** ownership has been studied most frequently. Hattori and Lanza [28] discussed code ownership as identifying the developer who owns an artifact of a software system "by measuring who has accumulated more knowledge of each artifact". Nagappan et al. [55] studied the impact of organizational factors on code quality. Among the eight organizational metrics, they invested in the Depth of Master Ownership (DMO). They defined the developer with more than 75% edits in a binary file as the owner and considered their organizational level as the DMO. They also defined the level of Organizational Code Ownership (OCO) as the percentage of edits from the organization containing the owner of the binary file. The authors also introduced Overall Organization Ownership (OOW) as "the ratio of the percentage of developers at the DMO level making edits to a binary relative to the total number of engineers editing the binary." Meng et al. [49] discussed line-level code authorship by introducing repository graphs, structural authorship, and weighted authorship. They used Structural authorship and weighted authorship to extract the development history of a line of code and approximate its authorship. “Structural authorship is a subgraph of the repository graph. The nodes consist of the commits that changed that line. Development dependencies between the subset commit form the edges." Weighted authorship "is a vector of author contribution weights derived from the structural authorship of the line. The weight of an author is defined by a code change measure between commits" [49]. Businge et al. [13] used the amount of code change as defined by earlier studies [7] to estimate the code ownership for an application. In the extreme programming (XP) methodology, the ideal code ownership scenario is defined as “any developer can refactor any area of code in an application as long as it continues to meet the contract defined by interfaces and unit tests" [7]. As the opposition, Judy et al. [35] considered the notion of code ownership, stating that a single accountable authority on such is impractical. Hence, they introduce *product owner* as the individual (one person and not a team) that is accountable for achieving business objectives. Overall, 33 Papers specifically referred to the code ownership artifacts.

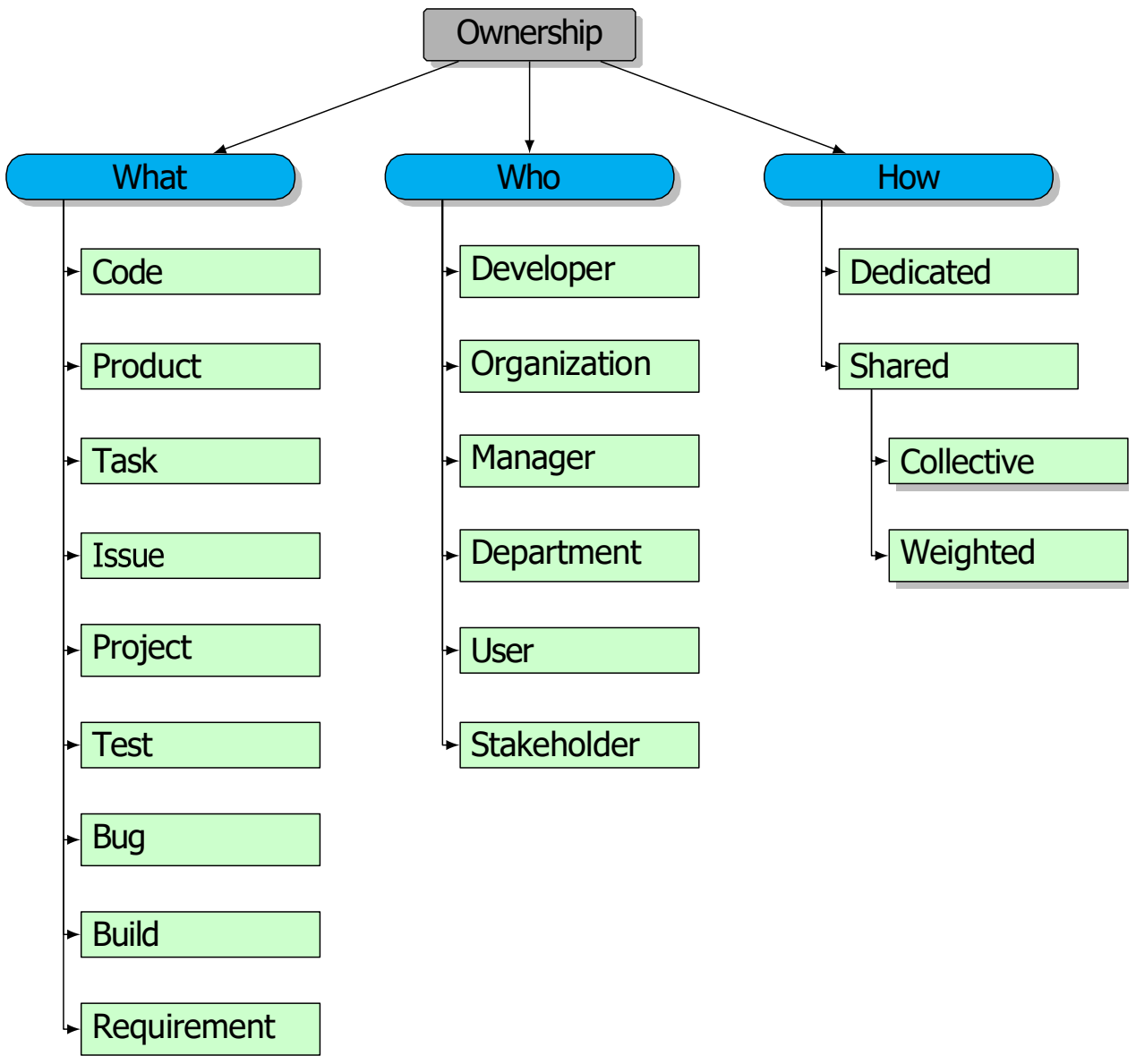

**Fig. 5** Taxonomy of the nature (What?), the owner (who?), and the degree of ownership (How?) as the result of our systematic review.

**"Product"** is the second most frequently appeared artifact in our literature review, and its ownership has been discussed in the literature, though only four studies specifically mentioned it. The product owner is a popular term for agile projects. Judy et al. [35] stated that the product owner is typically one person responsible for managing the product backlog, which contains an emerging set of requirements. They also suggest that the product's success depends largely on the decisions the product owner makes. Additionally, Shastri et al. [92] note that the product owner is often the customer representative. Judy et al. [35] state that the product owner is typically one person responsible for managing the product backlog, which contains an emerging set of requirements. The success of a product beyond the reliability, supportability, scalability, and time to market of its implementation entirely depends on the decision of the product owner [35]. Furthermore, Shastri et al. [92] note that the Product Owner is also commonly regarded as the customer representative.

**"Task"** ownership was specially studied for agile projects. In a team, team members take ownership of tasks from the task board by taking issues and risks [91]. Datta [17] examined factors that influence task ownership in Android development by introducing eight different variables.

**"Issue"** ownership artifact is discussed in two papers in our literature review. Caglayan et al. [14] discussed issue ownership by identifying the developer who resolves the issue. The authors showed that a small group of developers tends to take ownership of a large portion of new issues especially when the active issue count is relatively high in the software development life cycle. Issue ownership can be individual or teamwork [5], and the studies differentiate between the developer assigned to the issue and the one who resolves the issue [4].

**"Project"** ownership occurs when all team members take the responsibility to complete the project. It appears three times in our literature review, but one paper defined it clearly. In [19], the authors stated that in project ownership, all the team members are responsible for work completion, and they also share benefits among them. Burga et al. [12] discussed that both project managers and project have a relationship within the project governance structure.

**Table 3** Types and definition of corporal ownership in software engineering state of the art (**RQ1**).

| Artifact | Definition | Study |
|---|---|---|
| Code | Code author is the developer who writes a piece of source code. Code owner is defined by measuring who has accumulated more knowledge of each software component (e.g. module, file, class, binary). | S1, S3, S4, S5, S6, S7, S8, S10, S11, S12, S13, S15, S16, S17, S18, S19, S20, S21, S23, S30, S35, S36, S37, S39, S41, S42, S43, S44, S45, S49, S57, S62, S70 |
| Product | The product owner is responsible for managing the product backlog with the aim of maximizing the project's value. | S2, S48, S59, S69, S78 |
| Task | Task owner is the developer who accepts the responsibility of doing a task from a team's task list. | S14, S52, S64 |
| Issue | Issue owner is the developer who identified and resolved an issue in an artifact, and he may not be the same person who has worked on the artifact previously. | S9, S73 |
| Project | Project owners are team members collectively responsible for completing the work of a project. | S46 |
| Test | Test owner is any distinct developer that owns at least one test in the test suite. | S25 |
| Bug | Bug owner is the developer who has written the code that caused the bug. | S24 |
| Build | The build owner is the proportion of developers responsible for build maintenance. | S28 |
| Requirement | Requirement owner is a person who provides a requirement/requirements for a system. | S38 |

**"Test"** ownership has also been discussed by Herzig and Nagappan [30] and has later been referred to by studies focusing on code quality. Within a software team, a test owner is a developer who owns at least one test case from a test suite [30]. It appears only once [30] in our literature review where the authors showed that organizational structure impacts test case quality. They defined test case quality through test effectiveness and test execution reliability. They calculated the Spearman rank correlations between sixteen organizational metrics and measures of test suite effectiveness and test suite reliability and suggested that test suites whose owners are distributed over multiple organization subgroups with long communication paths are negatively correlated with quality.

Along the same line, **"Bug"** ownership was studied by Zhu et al. [102]. The developer who submitted the bug-introducing commit is known as the bug owner [102]. A bug can be fixed by the developer who introduced it or by a different developer. Zhu et al. [102] found that bug-fixing time by the bug owner is much shorter than that of other developers. They also found that the owners' bug-fixing commits are larger and have a different pattern from that of non-owners.

**"Build"** ownership is mainly discussed to maintain and change build systems. In their study, McIntosh et al. [47] proposed a definition of build ownership based on the proportion of developers responsible for maintaining a build system. The authors also identified two distinct styles of build ownership. The first style, referred to as "concentrated," involves a small dedicated team responsible for build system maintenance. The second style, called "dispersed," is characterized by the majority of developers contributing code to the build system. Shridhar et al. [93] studied six different build change categories and three different types of ownership styles.

**"Requirement"** is a newer concept among all artifacts of ownership. Hadad et al. [25] studied requirement ownership by identifying the ownership of requirements of a system. They studied five different types of such ownership where only developers are the owner of each requirement, where clients provide requirements of a system, where users are the author of a system requirement, or where ownership is collaborating and engineer, client, and user are equally responsible for every requirement, or where ownership is diffused. There is no one person for a requirement.

#### *4.1.2 Type of Owner (Who?)*

In our literature review, the majority of studies (82%) focus solely on identifying a set of "Developers" ($D_i$ where $i \geq 1$) as the owners of an artifact. In this context, "developer" is a general term used to refer to any member of the software team.

A subset of studies (14 out of 79) distinguishes between team members, either based on their organizational status or their experience within the team. For instance, Thongtanunam et al. [97] discussed reviewing activity in terms of code ownership. They introduce certain control variables using code reviews to identify code ownership. In their study, they expand the concept of code ownership to include review activities and consider the reviewer rather than just the developer. Similarly, Herzig et al. [30], test ownership is explored from the perspective of test engineers.

Nagappan et al. [55] investigated different ownership metrics at the **"Organizational"** level and analyzed the impact of organizational structure on software quality. In particular, they consider engineers, managers, and sub-managers within their analysis and introduce a number of metrics to capture the overall organizational ownership. (see Appendix I, Table 15, for further details). Bird et al. [8] examined the ownership effect on code quality. While ownership metrics are discussed here from the developers' perspective, **"Managers"** are also responsible for checking the code of a developer with less relevant experience. Greiler et al. [24] replicated [8] by introducing some more ownership metrics, which are about the organizational level of the developers and managers. Meneely et al. [48] conducted their analysis at the "department" level and found that team characteristics, such as team size and team expansion rate, affect software quality. Their main focus was on analyzing collaboration patterns among developers at the department level to determine whether a lack of inter-departmental collaboration impacts overall code quality. Hadad et al. [25] introduced requirement authorship and discussed five authorship types. They consider **Users**, **Stakeholders**, and **Developers** as the owners of software requirements in their research. In Figure 5, we abstracted these roles into six major entities by referring to the common roles within software teams.

#### *4.1.3 Degree of ownership (How?)*

The majority (60 out of 79) of the studies acknowledged and formulated the authorship issue as a problem of assigning a degree of responsibility to the different team members who edited a code snippet. For a software team consisting of $D$ individuals represented as $Developer = \{dev(1), dev(2), ...dev(i)\}$ and a set of artifacts developed by the team $Artifact = \{art(1), art(2), ..., art(j)\}$ our literature review and synthesis resulted in three general ownership models.

With the introduction of version control systems, software engineering literature gained the ability to study the concept of ownership and model fuzzy and shared ownership between multiple developers. The most commonly accepted ownership model is the **"Weighted"** model, which was followed by 31 out of the 79 studies in our systematic literature review. In this model, the ownership of an artifact is determined based on factors such as familiarity or effort spent on the artifact. The ownership of each artifact is represented as a vector of owners, where each owner is assigned a weight corresponding to their contribution to the artifact's changes. For example, Meng et al. [49] defined the weighted ownership of a code snippet as a vector of developers' contribution weights using the churn per commit. Where the total amount of contributions made by a developer $d$, to an artifact $a$ is $Contribution(i, a)$;

$$Weight(d, a) = \frac{Contribution(d, a)}{\sum_{i=1}^{i=D} Contribution(i, a)} \tag{1}$$

and

$$\sum_{i=1}^{i=D} Weight(a, i) = 1 \tag{2}$$

In this model, regardless of the unit used to measure a developer's contribution to an artifact (e.g., number of lines of code, churn, number of commits), the weight represents each developer's relative contribution to the overall team.

The literature, in particular, refers to two special cases of this model. First, where there is only one developer is identified as the owner of an artifact. Second, every developer on the team is responsible for all the artifacts, regardless of their degree of involvement. The problem of finding one developer as the responsible individual for writing and maintaining a line of code is commonly known as the "authorship" problem in software engineering [9]. **"Dedicated"** Ownership searches for one responsible developer for an artifact and identifies only one individual as the owner of an artifact. In this case, in Equation (1) for developer $dev(d)$, which is the author of an artifact;

$$Weight(a, i) = \begin{cases} 1, & \text{if } i = d \\ 0, & \text{otherwise} \end{cases} \tag{3}$$

Collective ownership focuses on completing tasks (such as bug fixes) by leveraging the entire team's expertise. Teams with a high level of collective ownership have all team members taking ownership of the software product. The assumption is that every team member has a minimum degree of expertise and knowledge of others' work through collaboration, coordination, or considering their organizational role. Maruping et al. [46] employ the notion of collective ownership.

In this case for any $art(a)$ and $dev(i)$;

$$Weight(a, i) \mathrel{!}= 0 \tag{4}$$

"Dedicated" and "Collective" ownership models represent the two extremes of the ownership spectrum where either an individual (dedicated) or the whole team (collective) is responsible for an artifact.

4.2 Ownership Modeling Treatments and Outcomes (RQ2) - What treatment and outcomes have been used in modeling ownership in software engineering literature?

Ownership models in software engineering use various attributes and artifacts. We followed the general experimentation process outlined by Wohlin et al. [99], as illustrated in Figure 2. As a result of the systematic review, we identified 18 unique dependent variables, 73 independent variables, and 18 control variables. Detailed descriptions of these variables can be found in Appendix II and Table 15. Table 4 presents only those dependent, independent, and control variables that were used more than once in our literature review. As a result of the systematic review, we identified 18 unique dependent variables, 73 independent variables, and 18 control variables. Details about these variables can be found in Appendix (II) and Table 15. Table 4 presents only those dependent, independent, and control variables used more than once in our literature review.

We identified 73 identical *independent variables*. The most popular independent variable in our literature review is the number of developers who changed a software component. The number of developers has been used as the independent variable in nine papers and in relevance to the ownership of code and issues. The other popular independent variable within our literature review is the number of major and minor contributors. These two were discussed in accordance with each other and within six papers. Here, a major contributor is a contributor who made more than a predefined threshold of the changes (commits) to a component. Similarly, a minor contributor is a contributor who made less than a predefined threshold of the changes (commits) to a component, and ownership is the highest value of the ratio of contributions performed by all developers. Code churn, code size, and the number of commits have been discussed in five to three studies. 63 independent variables are unique to one study within our literature review, which is mostly relevant to test ownership, code ownership, and product ownership. Another study within the scope of our studied literature has never chosen these independent variables. Details of all these variables are in Appendix (I).

**Table 4** Independent, Dependent, and controlled variables used "more than once" in modeling ownership studies.

| **Independent Variable** | **Study** |
|---|---|
| Number of developers changing a software component | S3, S6, S7, S9, S15, S16, S18, S19, S26 |
| Major contributor (developer with a majority of commits to a component) | S7, S15, S16, S19, S23, S29 |
| Minor contributor (developer with a minority of commits to a component) | S7, S15, S16, S19, S23, S29 |
| Highest % of commits a developer made to a software component | S7, S15, S16, S19, S23 |
| Code size | S19, S43, S54, S65 |
| Code Churn | S3, S4, S6, S19, S77 |
| Number of commits | S4, S11, S18, S77 |
| Code complexity | S3, S6, S77 |
| Developer experience | S8, S54 |
| Number of managers | S16, S25 |
| **Dependent Variable** | **Study** |
| Number of post-release failures | S3, S6, S7, S9, S11, S19 |
| Number of bugs | S13, S15, S16, S18, S23, S54 |
| Number of pre-release failures | S7, S11, S21, S29 |
| **Controlled Variable** | **Study** |
| Code complexity | S7, S11, S15, S23, S29, S72 |
| Code size | S7, S11, S15, S21, S29 |
| Project size | S1, S8, S23, S62 |
| Code churn | S7, S11, S21 |
| Developer experience | S1, S72 |
| Number of developers | S8, S21 |
| Number of developers participating in the review process (reviewers) | S14, S21 |

We identified a total of 18 *dependent variables*, which are detailed in Appendix I. The two most frequently appearing dependent variables in our literature review were the number of bugs and the number of post-release failures, each appearing in six studies. This was followed by the number of pre-release failures, which four studies used as a dependent variable. The remaining 15 dependent variables were used only in one study. Nearly one-fourth (23%) of the studies in our literature review selected some form of code quality metric as a dependent variable, identifying ten different metrics used in code ownership research.

We also identified 18 distinct *controlled variables*. The most popular controlled variable is code complexity, chosen by six different studies. Size, specifically code size, was chosen by five studies, and project size was selected by four different studies as a controlled variable. Additionally, code churn, developer's experience, number of developers, and number of code reviewers were used as controlled variables more than once. Our analysis revealed a total of 18 controlled variables, with 11 of them appearing in only one study within our literature (detailed in Appendix (I)).

### 4.3 Analytical Model, Replicability, and Evaluation setup (RQ3) - What analytics and research methods have been employed in the ownership studies, and to what extent are these studies replicable?

To summarize the papers in our systematic review, we examined their analytical models and categorized them using the four models introduced by Kaisler et al. [39]:

*Descriptive Analytics* is the most basic and commonly used model. It analyzes existing data statistically to identify patterns and understand the overall state of the data.

*Diagnostic Analytics* focuses on identifying the root causes of issues or anomalies within existing data. It answers the "why" question by providing deep insights into why certain events occurred.

*Predictive Analytics* utilizes machine learning techniques to forecast likely future outcomes based on existing data. This type of analysis helps to anticipate potential outcomes and enables informed decision-making.

*Prescriptive Analytics* combines descriptive and predictive models to guide the best course of action for achieving optimal outcomes. This model is focused on the decision-making process.

Variety of analytical models for software and project management exists in software engineering research [72] [77] [45] [87] [78] [66] [27] which we build our study upon such studies [88] [69] [58] [63] [59] [70] [60] [89].

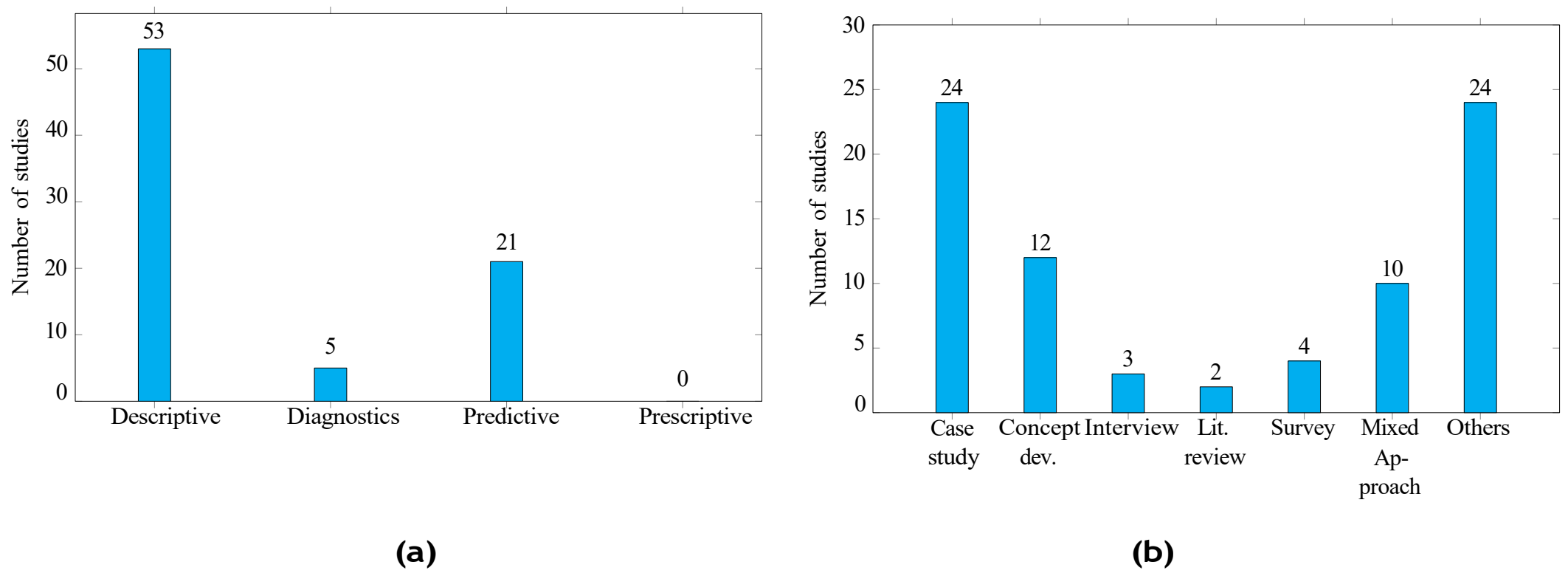

**Fig. 6** Distribution of paper based on (a) Type of analytics and (b) Research Method

**Table 5** Data availability within our systematic literature review. × shows the unavailability of the data while ✓ represents data availability. The table only includes papers with analytical components.

| Paper | Data | Dataset Description |
|---|---|---|
| S3 | × | 3,404 binaries exceeding 50 Million lines of code of Windows Vista. |
| S4 | × | 2,429 changes performed by four developers from one commercial project. |
| S5 | × | Organizational database (POST) to get the information of 29 to 252 developers having from 12 to 482 followers and the code changes they made over a period of seven years. |
| S6 | × | 511 lines of code for Windows Vista over seven teams in different geographical locations. |
| S7 | × | Commit histories and software failures of two Microsoft projects, Windows Vista and Windows 7. |
| S8 | × | Four open-source project Apache, Nautilus, Evolution and Gimp. |
| S9 | × | A large-scale enterprise software with more than twenty years of development history, and (ii) a smartphone Android operating system developed by Google and distributed as open source. |
| S10 | × | 10 open source software products written in Java. |
| S11 | × | Source code repositories, bugs, and the locations of developers of Firefox and Eclipse projects. |
| S12 | × | Student-submitted solutions to programming assignments written in C from RMIT University. |
| S13 | × | Five open source projects Dyninst, Httpd, GCC, Linux, GIMP. |
| S14 | ✓[1] | Android code review data from Gerrit-Miner. |
| S15 | ✓[2] | Seven open source projects Apache Ant, Camel, Log4J, Lucene, Eclipse JDT Core, PDE UI, and Equinox. |
| S16 | × | Four major Microsoft products being Office, Office365, Exchange, and Windows repository. |
| S17 | × | Change history from the version control system for Visual Studio 2013. |
| S18 | × | Repository of six open source Apache projects |
| S19 | ✓[3] | Repository of seven open source project |
| S20 | × | Four open source software Ant, Gremon, Struts2, Tomcat. |
| S21 | × | Code review dataset of two open source project Qt and Open Stack. |
| S23 | × | 45,000 open source Android-based repositories hosted on GitHub. |
| S24 | ✓[4] | SStuBs dataset which contains 10,231 bug fixes from the top 100 Java Maven projects. |
| S25 | × | organizational metrics datasets provided by the CODEMINE project. |
| S26 | × | Code logs, bugs, and historical records from the Human Resources department from Cisco projects. |
| S27 | ✓[5] | Code and bugs of five open source project Angular.JS, Ansible, Jenkins, JQuery, and Rails. |
| S28 | × | Build log of 18 Apache and Eclips open-source projects over a period of fourteen months. |
| S29 | × | Code and reiews of a large-scale commercial project. |
| S30 | × | 16 years entire history of the Apache Ant system. |
| S31 | ✓[6] | Pull Requests of 246 open source projects hosted on GitHub. |
| S33 | × | 41,140 open source code file changes from Eclipse and 6,548 function changes for a proprietary software project between January 2001 and September 2013. |
| S34 | × | 132 versions of PLAY released between 2008 and 2013 including 14 evolutionary and 118 maintenance releases. |
| S35 | ✓[7] | 270,000 commits of IntelliJ IDEA Community Edition project authored by 500 developers between 2004 to 2020. |
| S36 | × | Source code of nine different authors from the Codeisplanet website. |
| S37 | × | Four open source systems namely ArgoUML, Mozilla, Samba, and Squid. |
| S39 | × | 150 collaboratively written C++ files for 106 programmers active on GitHub. |
| S40 | × | GNU C compiler GCC version 3.3.2 open source project. |
| S41 | × | 1,000 programmers' data, collected from Google Code Jam (GCJ). |
| S43 | × | Six Apache projects naming Activemq, Aries, Carbondata, Cassandra, Derby, Mahout. |
| S44 | × | 200,000 source code files written by 898 developers from open source projects on GitHub. |
| S45 | × | Several open source projects, including Apache HTTP Server, Dyninst, GCC and other projects from GitHub. |
| S50 | × | SDT software project user requests over six months. |
| S51 | × | 200 open source projects belonging to Android, Apache, and Eclipse. |
| S53 | × | Development history of four releases of a company. |
| S54 | × | Commit history and issue from 12 GIT repository. |
| S57 | × | Google Code Jam (GCJ) development log between 2008 to 2016 and 1,987 repositories hosted on GitHub. |
| S62 | ✓[8] | 66 releases of 118 open source projects on GitHub. |
| S63 | × | Developer's communication data for three open source projects naming Gaim, eGroupWare, and Compiere. |
| S66 | × | Jira issues from four open source Java projects: Hibernate, JBoss, Mule and Spring. |
| S68 | ✓[9] | 200 most forked Java projects from GitHub. |
| S70 | × | contest code from 2008 to 2016 of Google code Jam |
| S72 | × | 1.2 million commits and more than 25,000 developers of multiple Open Source as well as a proprietary software project |
| S75 | × | Commits and issues of GitHub repository |
| S77 | × | Source code, version, and Jira issues of a commercial project |

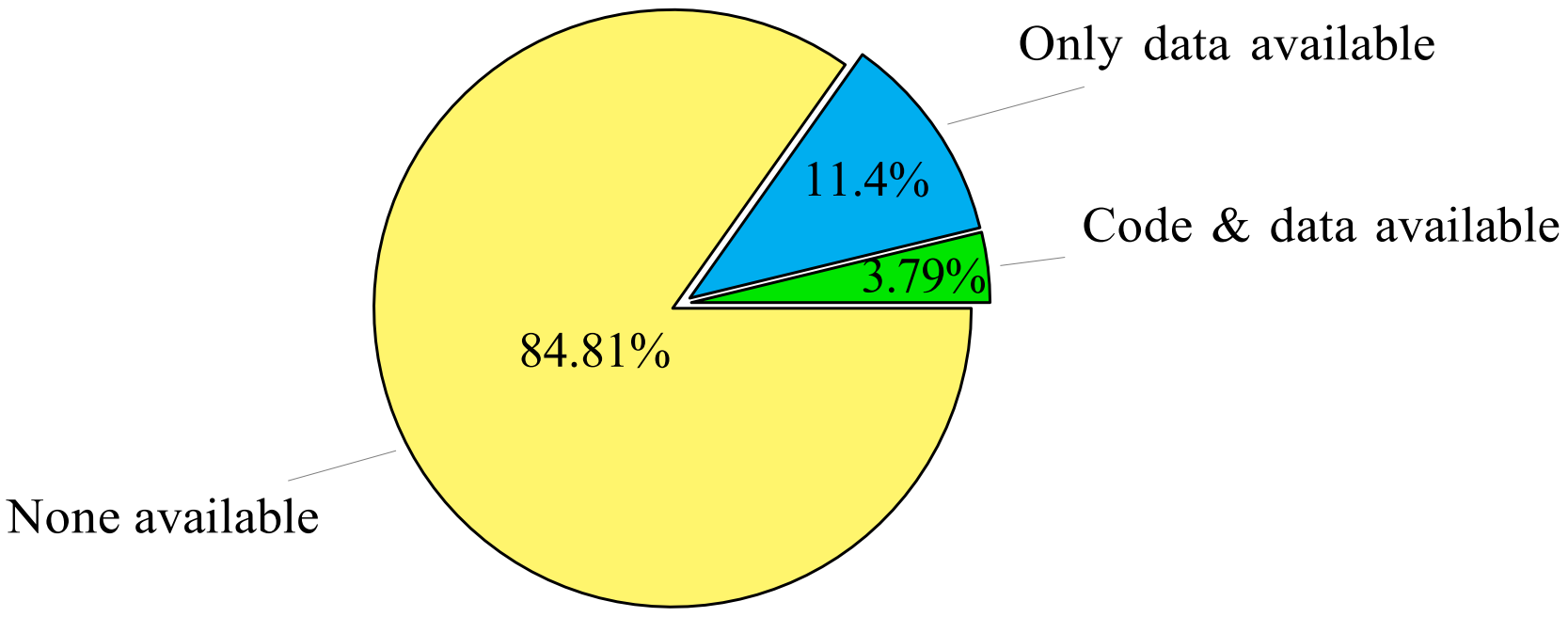

**Fig. 7** Percentage of papers for which code and data are available.

We applied Wendler's research method schema [98] to classify the type of research evaluation used in each study. The classification includes case studies, surveys, interviews, concept development, literature review, mixed approaches, and others. Additionally, we examined the availability of data and code for each study to facilitate replication, as per **RQ4**. The summary of our findings is presented in Figure 6-(a) and Figure 6-(b), where Figure 6-(a) displays the number of papers in each analytic category, while Figure 6-(b) illustrates the number of papers in each research method category. Descriptive analytics was the most commonly used analytic model employed in 67% of the studies, while we did not identify any studies using a prescriptive approach. Predictive analytics was employed in 21 studies, and four studies utilized diagnostic analytics. Figure 6-(a) provides a detailed breakdown of the analytics used in each study.

Regarding the research method, case studies were the most commonly used approach, with 24 studies following this method. Concept development was used in 12 studies, followed by interview-based and survey-based studies, which were used in three and four studies, respectively. Only two studies were based on literature reviews. Ten studies used a mixed approach, such as Drury-Grogan et al. [18], who performed both a case study and an interview to identify decision characteristics. Finally, 24 papers could not be classified into any of these research method categories and were tagged as "others." figure 6-(b) provides a detailed breakdown of the research method used in each study.

We were also interested in assessing the applicability status of these studies. We manually checked whether data or code was available for each study at the time of writing this paper (Figure 7). Only nine studies in our literature review provided a link to their dataset. Among them, three studies also included source code. Additionally, we observed that 32 studies utilized open-source data for their research. Table 5 provides an overview of these studies.

## 5 Benchmarking State of the Art Model in Brightsquid (RQ4) - How does the code ownership status in Brightsquid compare with the findings of three selected state-of-the-art studies?

The literature review was motivated by our partnered company's request to assess the status of code ownership in relation to the current state of the art or best practices. In the absence of any established benchmark, we conducted the systematic literature review (SLR) presented here.

## 5.1 Selecting Studies for Replication at Brightsquid

Our literature review revealed a substantial convergence in modeling code ownership and its associated variables, focusing on evaluating their influence on code quality. During our discussions with the partnered company, we adhered to two primary criteria for paper selection in the replication process at Brightsquid:

1. The study reports the impact of code ownership.
2. The study reports a case study evaluation of ownership of a priority software product or a comprehensive list of open-source products.

We identified two studies conducted at Microsoft. The initial study, authored by Bird et al. [8] [S7], is highly cited and frequently referenced in the literature. Subsequently, a follow-up replication and extension study was conducted four years later, once again at Microsoft, by Greiler et al. [24][S16]. We also decided to replicate the study by Foucault et al. [21][S19], which examined seven large-scale open-source repositories using a comprehensive set of metrics. These metrics have since been utilized in various other studies. We listed all the research questions of these studies In Table 6. We excluded two research questions, $RQ_{S7}3$ and $RQ_{S16}4$ as we did not have such data in Brightsquid. These two questions investigate the impact of the development process on the related outcome through qualitative studies. To answer each of the remaining research questions, we used the results of our systematic literature review, along with our extracted lists of dependent, independent, and controlled variables for each study. We then gathered relevant data and metrics to provide answers to these research questions.

## 5.2 Hypothesis and Research Methodology in Original Studies

While these studies have addressed nine research questions, there is a notable overlap in the hypotheses they tested, even though they employed different dependent, independent, and controlled variables. Bird et al. [8] formulated their research based on four hypotheses, two of which established a connection between software component ownership and the occurrence of failures:

**Hypothesis 1:** Software components with many Minor contributors will have more failures than software components with fewer.
**Hypothesis 2:** Software components with a high level of Ownership will have fewer failures than components with lower top ownership levels.

**Table 6** Research Questions addressed in the three replicated studies

| Study | Research Question | Answered? |
|---|---|---|
| Bird et al. [8] | $RQ_{S7}1$: Are higher levels of ownership associated with fewer defects? | Yes |
| | $RQ_{S7}2$: Is there a negative effect when a software entity is developed by many people with low ownership? | Yes |
| | $RQ_{S7}3$: Are these effects related to the development process used? | No |
| Greiler et al. [24] | $RQ_{S16}1$: Are ownership metrics indicative of code quality for other software systems than for Windows? | Yes |
| | $RQ_{S16}2$: For which levels of granularity (directory and source file) do ownership metrics correlate with code quality measured by the number of fixed bugs? | Yes |
| | $RQ_{S16}3$: Can ownership metrics be used to build classification models to identify defective directories and defective source files? | Yes |
| | $RQ_{S16}4$: What are the reasons for the lack of ownership? | No |
| Foucault et al. [21] | $RQ_{S19}1$: Does code ownership measured via the code ownership (CO) metrics, most valued owner (MVO), Minor, and Major, have a relationship with software modules quality, measured with their number of post-release bugs? | Yes |
| | $RQ_{S19}2$: If so, do these metrics provide an added value for predicting the number of post-release bugs of a software module? | Yes |

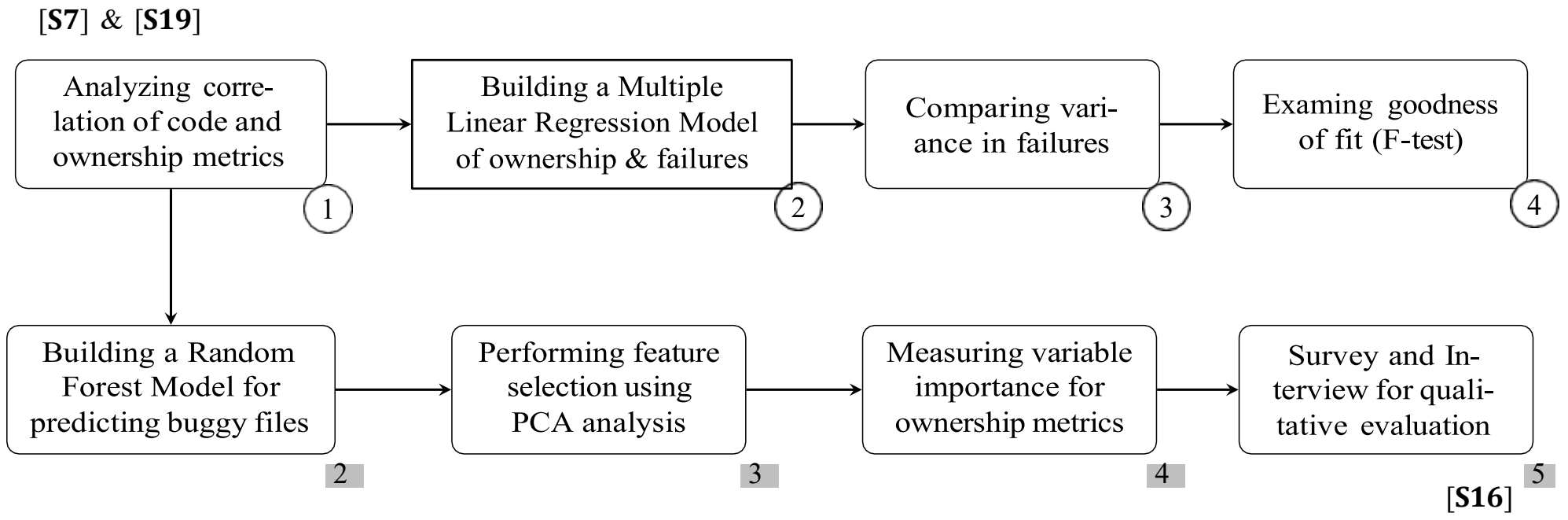

**Fig. 8** Main steps taken in the original studies of Bird et al. [8], Greiler et al. [24], and Foucault et al. [21] to analyze ownership. We refer to these steps in the text.

**Hypothesis 3:** Minor contributors to components will be Major contributors to other components that are related through dependency relationships.

**Hypothesis 4:** Removal of Minor contribution information from defect prediction techniques will decrease performance dramatically.

Analogous to this study, Greiler et al. [24] tested the same hypothesis with a few differences for the same corporation. They *altered the granularity* of their analysis by examining ownership at the file and code directory levels. Source directories contain source code files with the same enclosing path defined by the original study. Additionally, they *modified the ownership threshold for identifying* Minor *owners* to less than 50% ownership. The authors also introduced a new variable of *Minimal* owners, consisting of developers with less than 20% code ownership. Moreover, they separately analyzed files edited by only one developer, classifying them as having *strict ownership*.

Foucault et al. [21] also investigated the same four hypotheses but with several differences in their approach. They conducted their hypothesis testing across six open-source projects and used only the number of post-release bugs as their dependent variable. In identifying software modules, they relied on manual selection by the research team, defining a software module as either a file or a directory. Additionally, they *considered both the* Churn *metric and* Touche (the number of files touched by a developer) in their analysis. Furthermore, they introduced a new term for the Ownership metric, referring to it as Most Valued Owner (MVO). We summarized all the dependent and independent variables explained in Table 7 and separated for each of the three studies we replicated.

The empirical methods used to investigate these hypotheses by Bird et al. [8] and Foucault et al. [21] are largely the same despite differences in subjects and some metrics. In the first step (Step ① in Figure 8), they analyzed the correlation between code attributes and ownership metrics. Subsequently (Step ②), they developed a multiple linear regression model to identify the relationship between ownership metrics and the number of failures while controlling for source code characteristics. To test the fourth hypothesis and report the results, they compared the variance in failure (also known as $R^2$) (Step ③) and assessed goodness of fit using F-tests (Step ④).

Greiler et al. [24] followed a multi-step approach, similar to the other two studies (Step ①), by initially examining the correlation between variables. They then proceeded to the next step (Step 2 in Figure 8), where they constructed a Random Forest model to predict source code files and directories associated with at least one bug. In the subsequent step (Step 3), they conducted feature selection using Principal Component Analysis (PCA) to reduce the number of orthogonal dimensions. Step 4 involved measuring each ownership metric's variable importance by calculating the area under the ROC curve. To complement their quantitative analysis, they conducted interviews and surveys to qualitatively explore developers' perceptions in Step 5.

**Table 7** What, who, and How ownership was discussed in the selected studies for replication and the dependent, independent, and control variables.

| ID | What? | Who? | How? | Dependent variable | Control variable |
|---|---|---|---|---|---|
| S7 | Binary files (Code, Component) | Developer, Manager | Weighted | Number of pre-release failures, Number of post-release failures | Code size, Code churn, Code complexity |
| S16 | Binary files (Code, Component) | Developer, Manager, Organization | Collective shared | Number of bugs | N/A |
| S19 | Directories (Code, Component) | Developer | Shared | Number of post-release bugs | N/A |

**Independent Variables**

| ID | Metric name | Definition |
|---|---|---|
| S7 | Major | The number of developers who made more than 5% commits to a file |
| | Minor | The number of developers who made less than 5% commits to a file. |
| | Total | The total number of developers who contributed to a file. |
| | Ownership | Proportion of commits for the highest contributor to a file. |
| S16 | Minors | The number of developers who made less than 50% commits. |
| | Minimals | The number of developers who made less than 20% commits. |
| | Contributors | The Total number of contributors. |
| | Ownership | Proportion of commits for the highest contributor. This definition is identical to the one provided by Bird et al. [7] [S7] |
| | Avgownership | Average ownership values for all files in a directory: $(\sum fileownership)/(\#files)$. |
| | Ownershipdir | % of commits of the highest contributor considering all files in a directory: $(\sum ofmaincontributorcommits)/(\#allcommits)$. |
| | Minownerdir | % of commits of the lowest contributor considering all files in a directory: $(\sum ofcommitsofthelowestcontributor)/(\#allcommits)$. |
| | Avgcontributors | Average of distinct contributors among all files in a directory: $(\sum ofdistinctcontributorsperdirectory)/(files)$ |
| | Pcminors | % of contributors among all contributors with less than 50% commits for all directory files: $(\sum ofdistinctminors)/(\#contributors)$. |
| | Pcminimals | % of contributors among all contributors with less than 20% commits across all directory files: $(\sum ofdistinctminimals)/(\#contributors)$. |
| | Pcmajors | % of contributors among all contributors with more than or 50% commits across all directory files: in a directory: $(\sum ofdistinctcontributorswithmorethan50\%changes)/(\#contributors)$. |
| | Avgminimal | Average minimals in a directory: $(\sum ofminimalsperfile/\#files)$. |
| | Avgminors | Average minors in a directory: $(\sum ofminorsperfile/\#files)$. |
| | Minownedfile | The ownership value of the file with the lowest ownership value. |
| | Weakowneds | Number of files in a directory that have an ownership value of less than 50%: $(\#offileswith < 50\%ownership)/\#files$. |
| S19 | Major | The number of developers who made more than 5% commits to a file. |
| | Minor | The number of developers who made less than 5% commits to a file. |
| | NumDevs | The total number of developers who contributed to a file. |
| | Most valued owner (MVO) | Proportion of commits for the highest contributor to a file. |
| | Touches | The number of files touched by a developer. |

Empirical Data from Brightsquid

At the time of writing this paper, Brightsquid has a total of 39 projects, each starting at a different time between 2009 and 2023. A total of 52 developers have worked across all these projects. Also, 13,330 issues exist in the company's issue tracking system. For this study, we only considered the projects active after March 2018 for our analysis. As a result, we limited our analysis to nine projects at Brightsquid, as the others were inactive during that time. All the project source codes are hosted on GitHub, and the company uses Jira as the issue tracking system. The issues are being traced to the source code using automated integration tools. Below, we describe the data gathered for replicating the selected research questions from the studies, as listed in Table 6. We retrieved these data through GitHub and Jira APIs. A non-disclosure agreement governs the project; hence, all identifying information is anonymized. We summarized the project status with some conventional descriptive statistics and based on the number of commits, bug fixes, files, churn, complexity, and the lines of code in Table 8.

Bird et al. [7] examined ownership at the component level, defining components as a "unit of development that has some core functionality. [...] In Windows, a component is a compiled binary." We discussed the research questions (RQ) from each of the studies, S7, S16, and S19. We referred to each of these questions as $RQ_{S\#}$ in Table 6. After discussing with our industry partner, we considered the "source code file" as the component to replicate the analysis of ownership at the file level for answering $RQ_{S7}1$ and $RQ_{S7}2$. We retrieved the number of failures/bugs from the Brightsquid issue management repository on JIRA. By using integration tools between JIRA and GitHub, Brightsquid traces issues back to commits and code, allowing us to gather bugs linked to each source code file. Interestingly, we found that the majority of files have no bugs assigned.

To investigate directory-level ownership following the study by Greiler et al. [24], we aggregated the number of distinct bugs associated with all source files and used this information to infer the number of directory-level issues. However, unlike in the original study, we could not identify any files that had been edited only once. We found that all files in the selected Brightsquid projects had either been edited multiple times or not at all since their creation.

We then retrieved Independent variables as detailed in Table 7. Overall, we gathered three unique dependent, twenty independent, and three controlled variables from our chosen studies. These variables are listed in Table 7. Additionally, we gathered three code metrics – Size (measured by the number of lines of code in a file), Complexity (using cyclomatic complexity), and Churn (as the lines of code changed in each file) – from the source code repository to support our analysis. We could not obtain any information equivalent to "organization ownership" introduced by Greiler et al. [24]. As a result, our analysis was limited to individual ownership metrics, unlike the original study.

### 5.3 Empirical Results and Ownership Status Quo at Brightsquid

Figure 9 provides a summary of ownership metrics within the context of Brightsquid, based on measurements similar to those employed by Greiler et al. [24]. Our analysis of ownership statistics in the selected projects involved the use of a 50% threshold to distinguish between major and minor contributors. In Figure 9 - (d), it becomes evident that most contributors hold the status of minor contributors at the file level. In most projects, there is just one major contributor, while all others fall into the category of minor contributors. Notably, we also observed that the primary contributor is responsible for over 85% of the commits. At the directory level, we noticed that the average number of files per directory varies, ranging from a minimum of three in Proj1 to a maximum of 21 in Proj9. Most contributors fall into the category of minor contributors, making up more than 75% of the total. Furthermore, the average ownership percentage at the directory level consistently exceeds 72%. This suggests that the primary contributor was responsible for the majority of commits to both files and directories, indicating a strong sense of ownership at both levels.

**Table 8** Characteristics of the Brightsquid projects active as of March 2020 and included in our dataset.

| Project | Commits | Files | Bugfixes | LOC | Churn | Complexity | Developers |
|---|---|---|---|---|---|---|---|
| Proj1 | 69 | 8 | 11 | 981 | 930 | 34.1 | 5 |
| Proj2 | 178 | 84 | 210 | 4,509 | 2,726 | 29.2 | 2 |
| Proj3 | 213 | 67 | 74 | 21,741 | 3,184 | 2,628.2 | 5 |
| Proj4 | 12 | 6 | 8 | 123,873 | 109 | 513.8 | 9 |
| Proj5 | 422 | 54 | 105 | 30,161 | 18,555 | 5,509.1 | 5 |
| Proj6 | 80 | 11 | 18 | 1,331 | 510 | 232.3 | 7 |
| Proj7 | 555 | 42 | 53 | 45,528 | 13,578 | 271.1 | 6 |
| Proj8 | 1085 | 292 | 491 | 9,685 | 5,732 | 1,161.8 | 5 |
| Proj9 | 1350 | 339 | 740 | 50,918 | 37,046 | 579.4 | 9 |

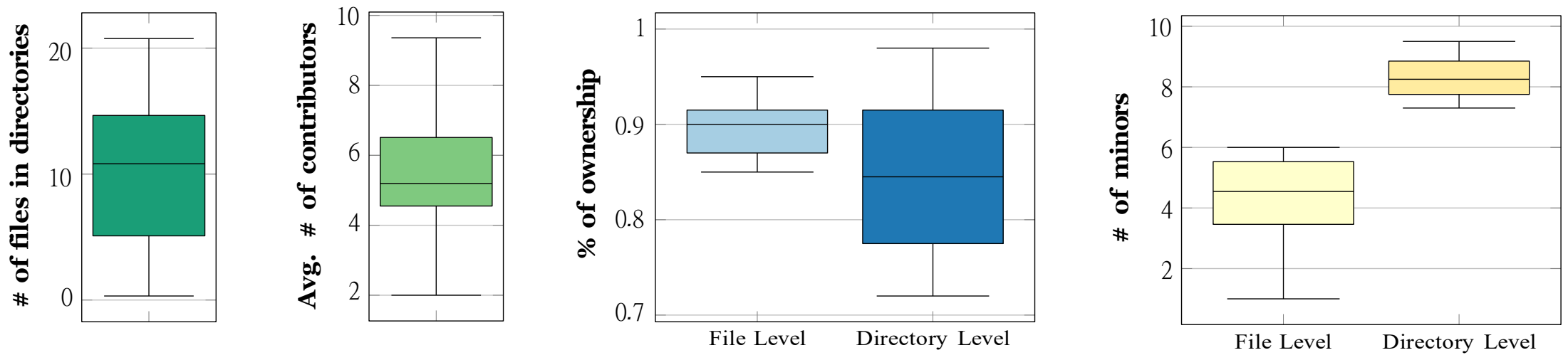

**(a) Code Directories** **(a) Code Files** **(c) Main Contributor's Ownership** **(d) Num. of Minor Contributors**

**Fig. 9** Descriptive statistics of ownership metrics of Brightsquid projects. The metrics were calculated based on Greiler et al. [24].

Figure 8 outlines the main steps of the methodology used in the three papers. We followed these steps using data from BrightSquid. When performing the analysis, we aggregated variables with different names but identical definitions, such as LOC, NumDevs, and MVO, which correspond to size, total, and ownership as defined by Bird et al. [8]. For each step, we followed the protocol of the original study.

### Step ① - The Correlation of Code Ownership Metrics and Quality:

We computed Spearman rank correlations between code ownership metrics and the number of bugs in each of our selected projects. We examined correlations between the number of bugs and all metrics at the file and directory levels. Specifically, we conducted a Spearman correlation analysis of the number of bugs with four ownership metrics: ownership, total, major, and minor, as well as three code metrics: code size, churn, and complexity. These analyses were performed for nine projects within Brightsquid, as detailed in Table 9.

*File Level Analysis:* Table 9 presents correlations at the file level. The findings show a significant correlation between the number of bugs and metrics such as ownership, major, and total across eight out of nine projects, with Proj2 being the exception. As depicted in Table 9, across all Brightsquid projects, we identified a negative correlation between ownership and the number of bugs. This suggests that higher levels of file ownership are associated with fewer bugs. In essence, when code ownership is fragmented among multiple developers (resulting in lower degrees of ownership), the likelihood of encountering bugs in the project increases. Additionally, we observed a positive correlation between the number of bugs and both total and major owners. Regarding the minor metric, we consistently observed a weak correlation across all projects, even when varying the minors threshold from 5% to 20% (referred to as minimals by Foucault et al. [21]), and then to 50% (as suggested by Greiler et al. [24]). Consistently, Minors did not show a strong and significant correlation. Similarly, in Brightsquid projects, we only observed sparse correlations between these metrics and the number of bugs and code metrics (see Table 9).

*Directory Level Analysis:* Table 10 displays Spearman correlation values for directory-level metrics. The correlation results were inconsistent across projects. We observed mostly weak correlations between ownership metrics at the directory level and the number of bugs across nearly all projects. Only Proj6 exhibited a significant correlation at the directory level. Furthermore, in five projects, the number of bugs decreased as the percentage of minor, minimal, and major contributors within a directory increased. However, in the remaining four Brightsquid projects, the opposite trend was observed, where the number of bugs increased with a higher percentage of these metrics.

While the directory-level results remained inconclusive, we found that the correlation between the number of bugs and ownership metrics was more significant at the file level.

**Table 9** Spearman correlation between metrics and the number of bugs on file level granularity. Bold and underlined values are absolute correlation coefficients above absolute values of 0.50 and 0.75, respectively

| Metric | Proj1 | Proj2 | Proj3 | Proj4 | Proj5 | Proj6 | Proj7 | Proj8 | Proj9 | Avg. |
|---|---|---|---|---|---|---|---|---|---|---|
| Ownership (mvo) | **-0.61** | -0.17 | **-0.72** | **-0.89** | **-0.75** | **-0.59** | -0.66 | **-0.66** | **-0.59** | **-0.63** |
| Major | **0.67** | 0.22 | **0.73** | **0.93** | **0.81** | **0.62** | **0.71** | **0.81** | **0.61** | **0.68** |
| Minors (< 5%) | -0.13 | 0.24 | 0.03 | 0.08 | 0.27 | 0.36 | 0.26 | 0.28 | 0.19 | 0.17 |
| Minors (< 50%) | -0.39 | **0.60** | 0.21 | **0.85** | **0.54** | **0.88** | **0.53** | **0.61** | 0.41 | 0.47 |
| Minimals | -0.21 | **0.5** | 0.2 | **0.82** | 0.45 | **0.82** | **0.52** | **0.58** | 0.41 | 0.45 |
| Total (NumDev) | **0.67** | 0.27 | **0.73** | **0.93** | **0.81** | **0.66** | **0.73** | **0.81** | **0.61** | **0.69** |
| Touches | 0.23 | 0.42 | 0.26 | -0.05 | 0.33 | 0.42 | 0.34 | 0.44 | 0.14 | 0.28 |
| **Code metric** | | | | | | | | | | |
| Churn | 0.26 | 0.29 | 0.24 | **0.74** | 0.32 | 0.43 | 0.23 | 0.48 | **0.54** | 0.39 |
| Size (LOC) | 0.41 | **0.52** | 0.19 | -0.22 | 0.4 | -0.01 | 0.28 | **0.7** | 0.4 | 0.30 |
| Complexity | **0.69** | 0.18 | 0.04 | **0.53** | 0.29 | 0.22 | 0.26 | 0.21 | 0.44 | 0.32 |

*In Brightsquid, stronger file ownership is associated with fewer* number of bugs *at the file level. Furthermore, the consistently strong correlation between the number of* total *and* major *owners is correlated with a higher* number of bugs.

**Step ② - Multiple Linear Regression Model of Ownership and Failure:** The correlation between the ownership and the code metrics motivated Bird et al. [8] to build regression models. It is essential to analyze if the increasing number of bugs in the projects is attributable to more minor contributors or to other measures such as size, churn, and complexity that are also known to be related to faults. Similarly, we used multiple linear regression to examine the relationship of ownership metrics while controlling source code attributes (here size, churn, and complexity). We built five statistical models for every project to examine how large or small the effect of each metric is on the number of bugs.

**Step ③, ④ - Comparing Variance in Failure ($R^2$) and Goodness of fit:** After constructing the five models, we evaluated and compared their variance in failure ($R^2$). We began with the Base model consisting of only the three code metrics: size, churn, and complexity. We subsequently added each ownership metric one by one to determine which variables influenced the increase or decrease of bugs, following the approach of Bird et al. [8].

We summarized the results of this analysis based on our regression model for Brightsquid in Table 11. In this table, the asterisk (*) indicates cases where we identified that adding a variable significantly improved the model through the goodness-of-fit F-test [8]. Among the five models, the base model comprises only three code metrics size, churn, and complexity. We observed that the Base model and the three code metrics primarily explained variance for Proj4 and Proj7, with percentages of 92.0% and 84.0%, respectively. By adding total and then minor as predictors, we observed that both variables significantly increased the proportion of variance for all projects except Proj5. On average, adding the minor variable increased the variance in failures by 6.7%, while adding the total variable increased this number by 7.63%.

**Table 10** Spearman correlation between metrics and bug numbers on directory level. Bold values are absolute correlation coefficients above 0.50.

| Metrics | Proj1 | Proj2 | Proj3 | Proj4 | Proj5 | Proj6 | Proj7 | Proj8 | Proj9 | Avg. |
|---|---|---|---|---|---|---|---|---|---|---|
| avgownership | -0.310 | -0.008 | -0.137 | -0.04 | -0.185 | **0.630** | -0.043 | 0.106 | 0.136 | 0.016 |
| ownershipdir | 0.188 | -0.328 | -0.127 | -0.047 | -0.069 | **0.579** | -0.213 | 0.075 | 0.137 | 0.022 |
| minownerdir | 0.145 | -0.328 | -0.132 | -0.042 | 0.0255 | 0.463 | -0.109 | 0.082 | 0.147 | 0.028 |
| pcminors | 0.218 | -0.352 | -0.121 | -0.048 | -0.080 | 0.802 | -0.157 | 0.084 | 0.143 | 0.054 |
| pcminimals | 0.306 | -0.352 | -0.125 | -0.047 | -0.130 | **0.617** | -0.226 | 0.077 | 0.136 | 0.028 |
| pcmajors | 0.218 | -0.352 | -0.134 | -0.042 | -0.047 | **0.802** | -0.142 | 0.022 | 0.150 | 0.052 |
| avgminors | 0.262 | 0.034 | -0.135 | -0.047 | -0.163 | **0.630** | -0.084 | 0.108 | 0.138 | 0.082 |
| minownedfile | **0.599** | 0.083 | **-0.66** | -0.049 | 0.062 | 0.011 | 0.063 | 0.242 | 0.135 | 0.054 |
| avgminimals | 0.320 | 0.034 | -0.136 | -0.04 | -0.136 | **0.630** | -0.136 | 0.107 | 0.138 | 0.087 |
| weakowneds | **0.623** | 0.344 | 0.102 | -0.024 | -0.051 | -0.037 | -0.088 | -0.132 | -0.201 | 0.059 |
| avgcontributo | **0.66** | **0.506** | 0.16 | -0.28 | **-0.571** | **0.626** | -0.08 | 0.281 | 0.255 | 0.173 |

**Table 11** Variance in failures for the base model (Code metrics), which includes standard metrics of complexity, size, and churn, as well as the models with Minor, Major and Ownership added

| Project | Base | Base+total | Base+minor | Base+minor+ major | Base+minor +major+ ownership |
|---|---|---|---|---|---|
| Proj1 | 38% | 38%(+0%) | 38.7%(+0.7%) | 38.9%(+0.2%) | 40.4%*(+1.5%) |
| Proj2 | 51% | 91%*(+40%) | 91%*(+40%) | 91%(+0%) | 91%(+0%) |
| Proj3 | 84.3% | 85.0%(+0.7%) | 85%(+0%) | 85.8%(+0.8%) | 85.8%(+0%) |
| Proj4 | 92% | 93.3%*(+1.3%) | 93.1%*(+1.1%) | 93.4%(+0.3%) | 97.6%*(+4.2%) |
| Proj5 | 56% | 56.04%(+0.4%) | 56.06%(+0.6%) | 57.4%(+0.8%) | 59.7%*(+2.3%) |
| Proj6 | 39.7% | 44.8%*(+5.1%) | 48.9%*(+9.2%) | 52.9%*(+4%) | 70%*(+18.1%) |
| Proj7 | 84% | 85.3%*(+1.3%) | 85.9%*(+1.9%) | 88.3%*(+2.4%) | 89.8%*(+1.5%) |
| Proj8 | 35.3% | 50%*(+14.7%) | 51.2%(+1.2%) | 51.2%(+0%) | 51.2%(+0%) |
| Proj9 | 42.5% | 47.7%*(+5.2%) | 48.1%*(+5.6%) | 49.7%*(+1.6%) | 49.9%(+0.2%) |
| **Avg.** | 58.09% | 65.68% | 66.44% | 69.62% | 69.48% |

We included major and ownership variables as predictors in the construction of the fourth (base + minor + major) and fifth (base + minor + major + ownership) models. However, we observed that these two metrics had a lesser impact than when we added the total and minor metrics, particularly in the case of Proj2 and Proj9. Consequently, we can conclude that, when controlling for code metrics, the number of total and minor contributors exhibits a strong relationship with the *number of bugs*. Furthermore, minor contributors have a greater influence on the increase in the *number of bugs* compared to the higher levels of ownership.
Our findings also show a positive correlation between the touches metric and the number of bugs, except for Proj4. However, this correlation is relatively weak, averaging at 0.21. When comparing the $R^2$ values between LOC and churn and the ownership metrics MVO, Major, and Minor, we observed that code metrics contribute significantly more to the total $R^2$ compared to the other metrics. Figure 10 illustrates and compares these results.

*In Brightsquid, the number of* total *and* minor *contributors shows a strong relationship with code quality when controlling for source code attributes. However, ownership metrics did not demonstrate strong predictive power compared to code metrics.*

**Step 2 , 3 - Random Forest Model for Predicting Buggy Files and Directories:** We built a Random Forest machine learning classifier to predict defective source files and directories. Following Greiler et al. [24], we divided our dataset into *files* and *directories* to investigate both granularities. We used two-thirds of one dataset for training and one-third for testing, along with 10-fold cross-validation. We also employed PCA (Step 3 - Figure 8) for feature selection and dimension reduction. Finally, we evaluated the classifier's precision, recall, and F1 score performance.

Table 12 summarizes the Random Forest classifier's performance in predicting defective sources at both the file and directory levels. At the directory level, we achieved an average F1 score of 0.84 (ranging from 0.61 to 0.93). At the file level, we obtained an average F1 score of 0.77 (ranging from 0.35 to 0.94). Our model performed relatively worse regarding precision and recall at the file and directory levels (except Proj1). This result indicates that ownership metrics can significantly affect code quality.

*The Random Forest model applied to Brightsquid showed potential for predictive modeling of code quality based on ownership values at both file and directory levels. However, the model performed poorly in predicting quality based on file ownership metrics for one of the nine projects.*

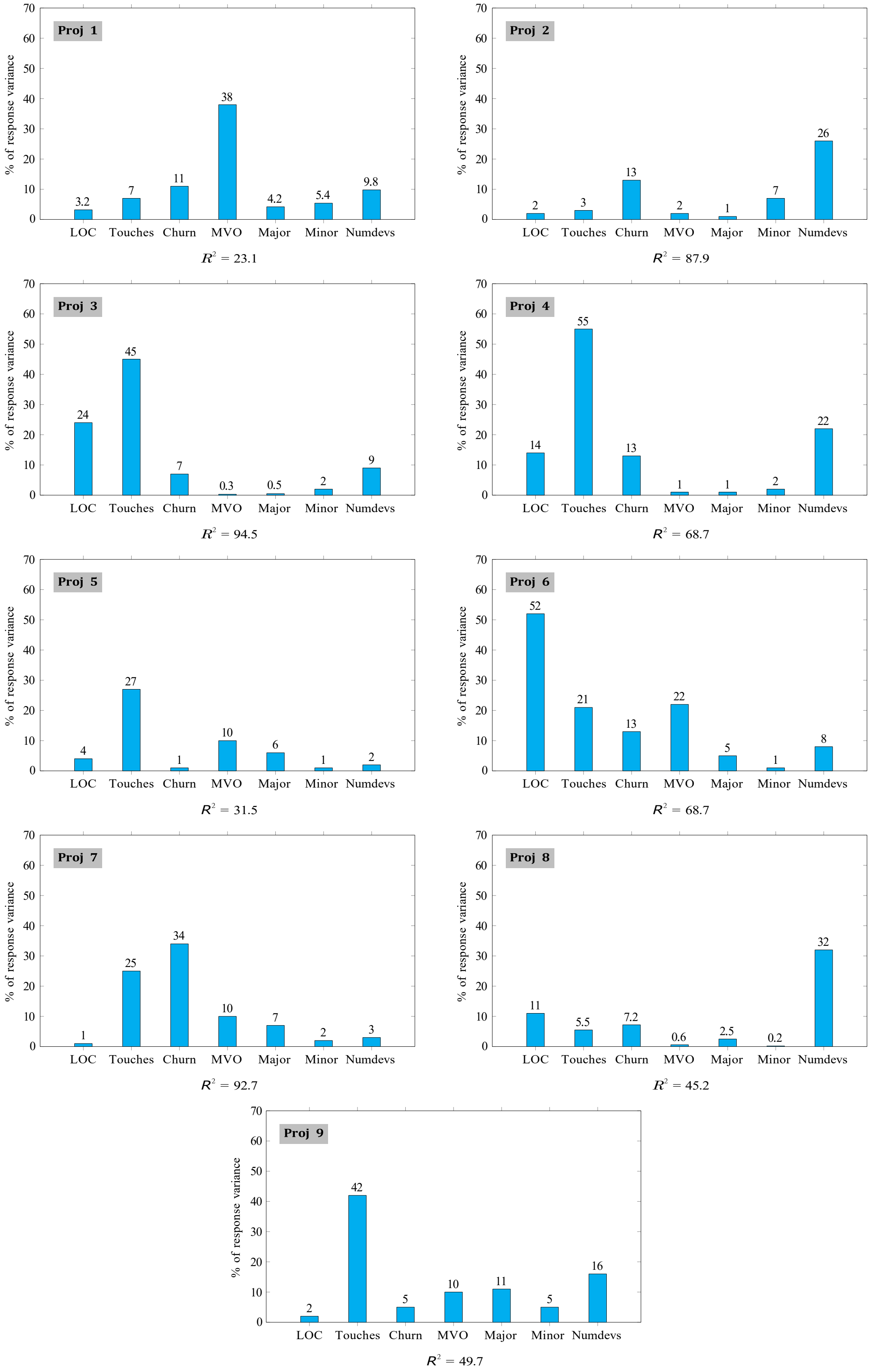

**Fig. 10** Relative importance of metrics in regression models using the number of bugs as the dependent variable.

**Table 12** Details on Precision, Recall, and F-measure for predicting defective source files and directories.

| | Directory level | | | File level | | |
|---|---|---|---|---|---|---|
| **Project** | **Precision** | **Recall** | **F1 score** | **Precision** | **Recall** | **F1 score** |
| Proj1 | 0.70 | 0.62 | 0.65 | 0.36 | 0.34 | 0.35 |
| Proj2 | 0.92 | 0.89 | 0.90 | 0.65 | 0.61 | 0.63 |
| Proj3 | 0.94 | 0.85 | 0.89 | 0.95 | 0.86 | 0.90 |
| Proj4 | 0.62 | 0.60 | 0.61 | 0.60 | 0.60 | 0.60 |
| Proj5 | 0.95 | 0.92 | 0.93 | 0.92 | 0.85 | 0.88 |
| Proj6 | 0.93 | 0.92 | 0.92 | 0.95 | 0.93 | 0.94 |
| Proj7 | 0.95 | 0.95 | 0.95 | 0.89 | 0.88 | 0.88 |
| Proj8 | 0.90 | 0.88 | 0.89 | 0.93 | 0.91 | 0.92 |
| Proj9 | 0.86 | 0.81 | 0.83 | 0.87 | 0.83 | 0.85 |
| **Avg.** | 0.86 | 0.82 | 0.84 | 0.79 | 0.76 | 0.77 |

**Step 4 - Variable Importance in Ownership:** Furthermore, we conducted a metric importance analysis for both the *file level* and *directory level* to assess the predictive power of the various metrics. To perform this analysis, we utilized the Random Forest model implemented in scikit-learn, specifically the Random Forest Regressor model. We present the metric importance scores for the *file level* in Table 13 and for the *directory level* in Table 14.

At the file level, Table 13 indicates that the most important metric for predicting defective source files is the number of contributors, except for Proj7 and Proj9. The least important metric for predicting defective source files varies by project. For example, in Proj2, Proj3, and Proj8, it is ownership, while in Proj1, Proj5, Proj7, and Proj9, it is minimals. Examining attribute importance at the directory level (see Table 14), minownedfile and avgcontributors consistently exhibit the highest importance scores for predicting defective source directories across all projects. Yet, the importance of other metrics varies across different projects. For example, in Proj3 and Proj8, all metrics except minownedfile and avgcontributors have low importance scores.

*At Brightsquid, the number of contributors to a file and the lowest ownership value among all the files in a directory exhibit the highest predictive power.*

## 5.4 Comparing Findings of Brightsquid Case Study and the Replicated Studies

To evaluate code ownership in BrightSquid, we replicated three studies and compared their findings with ours. In what follows, we solely compare our findings with the ones reported by the original studies. We marked differences between our findings and what was reported in the literature with Δ and tagged the similarities with ≈.

Δ <u>We did not find a strong correlation between minor contributors and number of Bugs:</u> In our first hypothesis, we tested whether components with many minor contributors tend to have more bugs than others. In Brightsquid, the correlation between the number of bugs and minor contributors was weak in all nine projects. This result differs from the findings of the original study [8][24][21], which suggested that software components with many minor contributors would have fewer failures.

**Table 13** Metric importance for prediction model based on ownership metrics classifying detective source files

| **Metrics** | **Proj1** | **Proj2** | **Proj3** | **Proj4** | **Proj5** | **Proj6** | **Proj7** | **Proj8** | **Proj9** | **Avg.** |
|---|---|---|---|---|---|---|---|---|---|---|
| ownership | 0.032 | 0.023 | 0.003 | 0.008 | 0.101 | 0.22 | 0.102 | 0.004 | 0.104 | 0.121 |
| minors | 0.109 | 0.078 | 0.007 | 0.007 | 0.061 | 0.112 | 0.078 | 0.008 | 0.116 | 0.064 |
| minimals | 0.013 | 0.112 | 0.005 | 0.026 | 0.014 | 0.112 | 0.023 | 0.012 | 0.051 | 0.041 |
| contributors | 0.121 | 0.268 | 0.095 | 0.221 | 0.022 | 0.085 | 0.034 | 0.087 | 0.165 | 0.122 |

**Table 14** Metric importance for prediction model based on ownership metrics classifying source directories

| Metrics | Proj1 | Proj2 | Proj3 | Proj4 | Proj5 | Proj6 | Proj7 | Proj8 | Proj9 | Avg. |
|---|---|---|---|---|---|---|---|---|---|---|
| avgcontributor | 0.14 | 0.161 | 0.145 | 0.33 | 0.162 | 0.27 | 0.101 | 0.142 | 0.131 | 0.176 |
| pcmajors | 0.109 | 0.077 | 0.007 | 0.007 | 0.061 | 0.112 | 0.078 | 0.009 | 0.116 | 0.064 |
| avgminors | 0.017 | 0.023 | 0.006 | 0.021 | 0.095 | 0.01 | 0.108 | 0.004 | 0.032 | 0.035 |
| weakowneds | 0.148 | 0.099 | 0.006 | 0.020 | 0.021 | 0.046 | 0.086 | 0.004 | 0.059 | 0.054 |
| pcminimals | 0.032 | 0.023 | 0.002 | 0.009 | 0.101 | 0.022 | 0.102 | 0.004 | 0.014 | 0.034 |
| pcminors | 0.013 | 0.112 | 0.005 | 0.027 | 0.01 | 0.112 | 0.023 | 0.011 | 0.051 | 0.040 |
| avgownership | 0.10 | 0.028 | 0.009 | 0.005 | 0.123 | 0.124 | 0.009 | 0.002 | 0.088 | 0.054 |
| minownerdir | 0.07 | 0.164 | 0.008 | 0.004 | 0.08 | 0.115 | 0.111 | 0.003 | 0.008 | 0.063 |
| ownershipdir | 0.10 | 0.09 | 0.006 | 0.006 | 0.018 | 0.104 | 0.023 | 0.001 | 0.068 | 0.046 |
| minownedfile | 0.211 | 0.184 | 0.195 | 0.176 | 0.201 | 0.197 | 0.22 | 0.205 | 0.164 | 0.195 |
| avgminimals | 0.107 | 0.018 | 0.005 | 0.001 | 0.102 | 0.085 | 0.178 | 0.011 | 0.015 | 0.058 |

Even when we varied the thresholds from 5% to 20% and 50%, the average correlation across the projects remained weak (i.e., ≤ 0.5), and the strength and direction of these correlations were inconclusive. In BrightSquid, we did not observe any significant correlation (Avg. = 0.17) between the number of bugs in software projects and minor contributors who had made contributions of less than or equal to 5%. However, the studies we reviewed indicated a strong correlation between minor contributors and the number of pre-and post-release failures. This finding contradicts the results reported in previous studies, which suggested a relationship between the number of bugs and Minor contributors.

*More number of* Minor *contributors does not necessarily lead to lower code quality. Unlike former studies, our case study at Brightsquid reveals the limitations of a direct correlation between these two metrics.*

≈ <u>We observed a strong correlation between ownership and total metrics and the code quality.</u> Previous studies have tested the hypothesis that software components with a high level of ownership have fewer failures compared to others. These studies showed a strong correlation between ownership and the number of bugs. Consistent with these findings, we observed a negative correlation with an average of −0.63 between the ownership metric and the number of failures in Brightsquid. We also found a strong correlation between total and the number of bugs, similar to previous studies results.

∆ <u>We observed a strong correlation between major contributors and number of bugs</u>. The studies by Bird et al. [8] and Foucault et al. [22] indicated a weak correlation between major contributors and the number of bugs. However, in our analysis of Brightsquid projects, we discovered a strong correlation (Avg. = 0.68) between major contributors and the number of bugs, independent of the size, churn, and complexity (i.e. the code metrics).

∆ <u>We did not find a strong correlation between the Total number of developers making changes to a file and the number of bugs.</u> While Foucault et al. [21] reported a significant correlation between the number of developers who modified a file (referred to as touches) and the number of bugs. However, in our analysis of the Brightsquid projects, we found no significant correlation between the number of file modifications (i.e., touches) and the number of bugs (Avg. = 0.28).

*In line with previous studies, high* ownership *demonstrated a strong correlation with improved code quality. In contrast to earlier research,* major *developers exhibited a notable correlation with the number of bugs. Moreover, contradicting prior literature, the* total *number of developers did not show a significant correlation with the number of bugs.*

∆ <u>We did not observe a strong correlation between code metrics and code quality at Brightsquid.</u> Bird et al. [8] reported a relatively strong correlation between classic code metrics (size, churn, complexity) and the number of bugs. However, we only observed a sparse correlation between code metrics and the number of bugs in BrightSquid projects. Previous studies tended to find a stronger correlation between code ownership metrics.

∆ We found no substantial increase in predictive power when using ownership metrics as opposed to code metrics. Bird et al. [8] hypothesized that the removal of minor contribution metrics decreases defect prediction model performance. Hence, they built a regression model and reported a statistically significant impact of code metrics on the number of failures. They reported a clear trend of a statistically significant relationship between ownership metrics and the number of failures in Microsoft Windows. In Brightsquid, our analysis showed a strong correlation between the number of total and minor contributors and code quality, even when controlling for source code attributes. However, we did not observe significant predictive power for ownership metrics compared to code metrics.

≈ The total number of developers is the most important factor for classifying defective source files. Foucault et al. [21], and Greiler et al. [24] assessed the significance of metrics for categorizing faulty source files and directories. Upon comparing our findings with theirs, we observed that the metric importance score for Brightsquid was very low. Nevertheless, similar to their studies, we discovered that the most important metric for classifying defective source files was contributors.

*Brightsquid, unlike previous studies, neither code metrics nor ownership metrics show statistically significant correlations with the number of bugs. Consistent with the existing literature, the* Total *number of contributors emerges as the most significant factor in identifying faulty files.*

∆ At the directory level, the lowest ownership value among all the files in a directory (minownerdir) has the strongest predictive power for bugs. When comparing our findings with the research of Greiler et al. [24] to assess metric importance in identifying faulty directories, minownerdir exhibited the highest significance in Microsoft products. However, the directory-level analysis in Brightsquid yielded less promising results. In contrast, for Brightsquid, minownedfile (representing the lowest ownership value among all the files in a directory) achieved the highest predictive power.

∆ We did not observe a significant correlation between directory-level metrics and the number of bugs. Greiler et al. [24] introduced 12 new ownership metrics at the directory level and reported a robust correlation between the number of bugs and the percentage of edits made by minor, minimal, or major contributors (pcminor, pcminimal, pcmajors, minownerdir). They discovered that an increase in minimal and minor contributors in a directory was associated with higher bug counts, while a greater presence of major contributors reduced the number of bugs.

In contrast, our analysis of Brightsquid showed stronger correlations between ownership metrics at the *file level* than those defined at the *directory level*. Specifically, we did not observe any significant correlation between any of the *directory-level* metrics and the number of bugs in Brightsquid. Therefore, we cannot draw definitive conclusions about the relationship between the number of bugs and the presence of minor, minimal, or major contributors in a directory.

*Unlike previous studies, directory-level metrics did not exhibit a significant correlation with the* number of bugs. *Moreover, the metric importance score for predicting faulty directories is very low in Brightsquid.*

We further discuss the implications of our study and these differences in the next sections.

## 6 Discussion and Implication of Findings at Brightsquid

In this section, we briefly summarize the implications of our studies, reflecting in particular on the differences we observed with the state of the art and the implications for research and practice.

### 6.1 Implications of Our Systematic Literature Review

We conducted a Systematic Literature Review (SLR) focused on ownership in software engineering. We categorized the studies based on the type of ownership artifacts examined. Code emerged as the most frequently discussed topic in the literature among the ownership artifacts. The literature review was motivated by the lack of any benchmark in the well-established field of software ownership, which made it impossible to make comprehensive comparisons. In this section, we present and discuss our findings, emphasizing their implications. Notably, we will highlight three main implications derived from this comprehensive literature review.

1. The software engineering community and researchers can utilize the results of our review to standardize the language and terminology employed in modeling ownership [46][51][26]. The terminology used by Bird et al. [8] has gained widespread adoption. However, several studies make use of the same variables. They frequently employ distinct terminology or repurpose variable names. This inconsistency creates confusion and impedes the clarity of modeling, thereby obstructing meaningful comparisons. In Appendix II of this paper, we provided a comprehensive list of dependent, independent, and control variables extracted from the 79 studies within our literature review. This divergence in terminology leads to a lack of synergy in leveraging each other's findings, resulting in considerable overlap within the literature analysis. Surprisingly, despite this overlap, the existing body of literature seldom reports direct comparisons. Our systematic literature review offers a consolidated collection of ownership models and variables (refer to Table 4), serving as a comprehensive resource to navigate the field of code ownership and facilitate the advancement of existing models.
2. Authorship refers to the process of determining who wrote a piece of software based on its source code. Traditionally, ownership of an artifact has been attributed to an individual (dedicated ownership), which is synonymous with authorship. However, with the advent of version control systems, shared ownership of artifacts is being represented either as a weighted value among team members or as a collective value among all team members. Code ownership and code authorship should be distinguished in terms of the method of measuring ownership.
3. While the research community strongly promotes the replicability of empirical research [36][2], the outlook is not very promising in the context of ownership, as we assessed in our systematic literature review. Among the ownership studies identified in our SLR, 84.8% are not replicable, while 67.1% of them are descriptive, with the majority presenting case study evaluations. (see Figure 6). We also listed and linked all the available repositories and associated case studies and their subjects in our review.

### 6.2 Implications Drawn from Our Case Study Findings

Brightsquid wanted to assess the status of code ownership in its repositories and compare it with other organizations. Our systematic literature review led to two studies on proprietary software, both conducted at Microsoft [8, 24]. Bird et al. [8] examined the relationship between different ownership measures and software failures.

When compared to the literature, our findings showed three main differences with the three studies we replicated, as summarized in Section 5.5. These findings should be discussed in terms of which software engineering activities take place.

**Firstly**, an increased number of Minor contributors does not necessarily lead to lower code quality. Depending on the development process and the business focus, it is sometimes unavoidable to have many developers working on the same code. Brightsquid prioritizes developers' accountability toward their customers over strict code quality. Therefore, when they receive and triage a customer request [64], Brightsquid assign it to a developer familiar with the customer and the received request. The assigned developer is then responsible for making decisions, individually or in collaboration with others, to accommodate the request. This approach is a key reason for having multiple minor developers per file or directory. The company believes that understanding customer needs and the context of their requests significantly impacts successful development. This strategy for assigning customer requests has shifted the dynamics of code ownership in Brightsquid. The company focuses on user stories and requests rather than simply assigning development tasks to developers, resulting in a significant number of minor developers working on a file/directory. Brightsquid is fundamentally oriented towards new market segments and innovative solutions. The development methods at Brightsquid are more adaptive and experimentation-driven, in contrast to the processes of the Microsoft Windows team, which focus on optimizing an existing product within an established market.

**Secondly,** our results showed that the increase in the number of major contributors is associated with more number of bugs in Brightsquid. This is while the former studies showed that the number of major contributors negatively correlates with code bugginess. It is essential to consider the context of the software where Bird et al. [8] and Grailer et al. [24] performed analysis on Microsoft products, including Windows. These products have a core focus that is less user-driven and primarily serve as engineering solutions for enterprises. In contrast, in Brightsquid, understanding the context of the user and their requests plays a pivotal role in successfully implementing changes or new features. In Brightsquid, despite maintaining a flat team structure, major contributors have limited familiarity and contact with individual customers. These major contributors are primarily associated with more volatile features within Brightsquid, focusing their major contributions on the core functionality of the system. These features are often more extensively tested and used; therefore, the proportion of bugs should be interpreted in this context.

**Thirdly**, Our findings indicated comparable outcomes regarding the correlation between robust file ownership and a reduced number of bugs at the file level, along with the correlation between the count of major contributors and a higher number of bugs. These results show that the number of developers contributing to a file (contributors) impacts the number of failures. However, unlike prior research, our observations did not reveal a substantial predictive capacity for code ownership metrics in forecasting code quality. While we could not identify the percentage of contribution as a conclusive factor, our results imply that Brightsquid should seek a balance between the expertise of the developers in maintaining particular parts of code (files and directories) and their familiarity with the user problems.

Traditional approaches for conducting software engineering research focused on producing generalizable results may not always be the most effective approach [10]. In addition to this, it's essential that we pay close attention to construct validity in empirical studies that delve into mining software repositories [84]. The outcomes of a case study, whether it's the one we've conducted at Brightsquid or those documented in existing literature, are inherently prone to this concern. Ralph and Tempero [84] pointed out that due to the inherent complexity in evaluating these measures, many empirical studies tend to skip assessing construct validity. It becomes increasingly clear that many studies have traditionally defined and modeled ownership by examining measures relative to the proportion of developers' contributions to the source code. However, the disparities between our findings and those of previous studies suggest that other factors, such as the development process, request prioritization, the skills of team members, and the quality and extent of testing, may exert a more significant influence on the occurrence of bugs in the software than simply the proportional contributions of developers. Furthermore, it's worth noting that our study at Brightsquid took place 12 years after the work of Bird et al. [8] at Microsoft and the open source projects analyzed. Consequently, the methodologies used in those earlier studies may not be directly transferable to the unique software development process or projects undertaken by Brightsquid. As put by Menzies et al. [50], simple global rules are inadequate for managing complex elements such as defect prediction and effort estimation.

Moving forward, we plan to empirically evaluate the impact of development processes on code ownership in Brightsquid as put by Brightsquid's CTO:

*"Looking into the results, I believe we should further invest in interface and unit test "contract" diligence as a greater predictor of software defects (at least in Brightsquid), in addition to the variables identified in the paper. Diligently defined, maintained, and owned test cases/suites (by developers, not by separate testers) are predictive indicators of quality. With greater freedom (co-design/co-development) comes greater responsibilities (i.e., diligence to well-defined contracts). To me, further analyzing test case ownership fits into the construct of strong governance, which perhaps needs to be in place to enable co-design/co-development. So, the more adaptive and evolutionary your solution is, the more important the maintenance of these test cases becomes."*

We believe that while there is a rich literature about code ownership, the field should further evaluate the impact of development and testing processes on these matters. We have established a long-term collaboration with Brightsquid [1] [62] [68] [29] [38]. Many of these collaborations were focused on release decisions and requirements management [76] [57] [71] [65] [75] [67] [61] [74] [73]. However, motivated by this study, we have performed a comprehensive survey and mining study for evaluating the role of developers' accountability and its impact on code ownership [44].

## 7 Threats to Validity

Threat to validity refers to the limitations or potential biases in research that can affect the validity and reliability of the results. In our study, we relied on a comprehensive classification by Wohlin et al. [99] to effectively identify and categorize various aspects of threats to validity.

**Are we measuring the right thing?** Construct validity in systematic literature reviews (SLR) refers to the degree to which the review's operational definitions of the key concepts or constructs being studied are accurate and complete. When performing a systematic literature review, using correct search keywords plays a critical role in identifying relevant studies. We identified 17 relevant search strings, which is a relatively high number compared to SLRs in software engineering. Furthermore, the manual analysis of numerous papers and the potential for human errors represent additional construct validity concerns. We ensured that at least two authors independently conducted each classification to mitigate this risk.

When it comes to mining BrightSquid data, we mostly followed the approaches of the three papers introduced and the measurements suggested by these studies. Multiple studies have tested the majority of the measurements and variables, yet only one of the studies has proposed a few (such as 'Touches'). Despite this, we independently evaluated the potential relationship between these variables and software ownership in BrightSquid. Nonetheless, we recognize that there were certain factors we did not examine, such as the potential influence of developer experience and education on the number of bugs. We acknowledge that these variables may have an impact on software quality and could potentially confound our results. However, our primary goal was to compare the status of Brightsquid using the state of the art and practice methods.

**Are we drawing the right conclusions about treatment and outcome relation?** Conclusion validity in a systematic literature review refers to the extent to which the conclusions drawn from the review accurately represent the evidence found in the studies included. This concerns whether the evidence supports the conclusions and whether alternative explanations for the findings have been ruled out. When performing the SLR, we formed our research questions to identify the *who*, *when*, *how*, and *what* software ownership has been discussed. We also identified different modeling aspects by systematically retrieving and unifying the dependent, independent, and controlled variables. Yet, there might be aspects that we missed in this comparison. We also analyzed all the active projects of Brightsquid at the time of this study. There were cases in which projects showed slight differences in their behavior (for example, correlation between the variables). These can be attributed to several socio-technical aspects and project staffing (for instance, human resource retention), which were not evaluated nor quantified in our study.

**Can the results be generalized beyond the scope of this study?** External validity in systematic literature review refers to the generalizability of the findings to the larger population outside of the studies included in the review. It assesses the extent to which the results can be applied to other settings, populations, or contexts. Our study was limited in scope as we only examined active projects from our partner company. This means that our findings may not be applicable to all the incoming projects or the projects done in other organizations, particularly open-source projects. Additionally, the development process for different projects can vary. Since all the Brightsquid projects analyzed in our study are written in Java, the conclusions and findings we present may not directly apply to projects developed in other programming languages because each programming language has its own unique programming paradigms, syntax, and libraries, impacting the development and testing process, tools and technologies which can influence software development process [6].

**Can we be sure that the treatment indeed caused the outcome?** For an SLR, Internal validity refers to the extent to which the review methodology was applied rigorously and the study results are accurate and reliable. We chose a comprehensive set of databases to search in our systematic literature review. However, there is a risk that these engines will not index other publications. Similarly, despite using 17 different search strings, the success of this literature review highly depends on the correct choice of keywords. When analyzing Brightsquid data, we built on the hypothesis that there is one developer as an owner of a project. Yet, this was mostly adopted from open-source projects. In Brightsquid, we observe that most developers are minor contributors across different projects. Moreover, Brightsquid does not maintain the organizational metrics, and the flat and fluid structure of the teams makes these factors irrelevant to this study. Hence, we only measured individual-level ownership and not team-level ownership. The absence of organizational metrics leaves room for speculation and interpretation, which could affect the accuracy of the observation.

## 8 Conclusion

Researchers have introduced various models to comprehend code ownership and its effects on developer performance metrics and code quality. Yet, systematic studies providing a comprehensive overview and taxonomy of these models have been lacking. Most of these models have been assessed in the context of open-source repositories, leaving a notable gap in our understanding of code ownership in proprietary software. To address this, we conducted a Systematic Literature Review (SLR) and replicated three experiments across nine proprietary projects within a software company.

Our study unveiled nine ownership artifacts, with code ownership being the most prevalent. This underscores the need for a standardized language and terminology in ownership modeling, as we identified numerous variables with similar meanings but different labels. Additionally, our examination of replicability in ownership studies yielded disheartening results, with a substantial 84.8% of the studies from our SLR lacking replicability. Moreover, most ownership studies rely on descriptive approaches and present case study-based evaluations. In our collaboration with Brightsquid, we identified three pertinent research papers on code ownership. After replicating these studies and adapting our assessment to Brightsquid's context, we found that, in contrast to state-of-the-art studies on code ownership, there is no significant correlation between the number of minor owners and code ownership status at Brightsquid. Nevertheless, similar to previous research, the number of contributors significantly impacts source code bugginess. Our study serves as a starting point for future research in this domain, aiding researchers in selecting appropriate comparisons and terminology.

Future work in this area could focus on developing more standardized methods for studying code ownership in proprietary software, exploring the relationship between ownership and other performance metrics such as developer experience, education, etc., and investigating how ownership models can be integrated into software development processes. Additionally, given that a mere 3.79% of ownership studies are replicable through code and data availability, this field requires more rigorous research methods and standardization to establish reliable benchmarks.

## Acknowledgment

This research was partially supported by ???.

## Appendix (I) Selected studies

[S1] Maruping, Likoebe M., Xiaojun Zhang, and Viswanath Venkatesh. "Role of collective ownership and coding standards in coordinating expertise in software project teams." European Journal of Information Systems 18.4 (2009): 355-371.

[S2] Judy, Ken H., and Ilio Krumins-Beens. "Great scrums need great product owners: Unbounded collaboration and collective product ownership." Proceedings of the 41st Annual Hawaii International Conference on System Sciences (HICSS 2008). IEEE, 2008.

[S3] Nagappan, Nachiappan, Brendan Murphy, and Victor Basili. "The influence of organizational structure on software quality." 2008 ACM/IEEE 30th International Conference on Software Engineering. IEEE, 2008.

[S4] Hattori, Lile, and Michele Lanza. "Mining the history of synchronous changes to refine code ownership." 2009 6th ieee international working conference on mining software repositories. IEEE, 2009.

[S5] Mockus, Audris. "Succession: Measuring transfer of code and developer productivity." 2009 IEEE 31st International Conference on Software Engineering. IEEE, 2009.

[S6] Bird, Christian, et al. "Does distributed development affect software quality? an empirical case study of windows vista." 2009 IEEE 31st International Conference on Software Engineering. IEEE, 2009.

[S7] Bird, Christian, et al. "Don't touch my code! Examining the effects of ownership on software quality." Proceedings of the 19th ACM SIGSOFT symposium and the 13th European. conference on Foundations of software engineering. 2011.

[S8] Rahman, Foyzur, and Premkumar Devanbu. "Ownership, experience and defects: a fine-grained study of authorship." Proceedings of the 33rd International Conference on Software Engineering. 2011.

[S9] Caglayan, Bora, and Ayse Bener. "Issue ownership activity in two large software projects." ACM SIGSOFT Software Engineering Notes 37.6 (2012): 1-7.

[S10] Corley, Christopher S., Elizabeth A. Kammer, and Nicholas A. Kraft. "Modeling the ownership of source code topics." 2012 20th IEEE International Conference on Program Comprehension (ICPC). IEEE, 2012.

[S11] Bird, Christian, and Nachiappan Nagappan. "Who? where? what? examining distributed development in two large open source projects." 2012 9th IEEE Working Conference on Mining Software Repositories (MSR). IEEE, 2012.

[S12] Tennyson, Matthew F. "A replicated comparative study of source code authorship attribution." 2013 3rd International Workshop on Replication in Empirical Software Engineering Research. IEEE, 2013.

[S13] Meng, Xiaozhu, et al. "Mining software repositories for accurate authorship." 2013 IEEE international conference on software maintenance. IEEE, 2013.

[S14] Datta, Subhajit. "Perspectives on task ownership in mobile operating system development (invited talk)." Proceedings of the 2nd International Workshop on Software Development Lifecycle for Mobile. 2014.

[S15] Foucault, Matthieu, Jean-Rémy Falleri, and Xavier Blanc. "Code ownership in open-source software." Proceedings of the 18th International Conference on Evaluation and Assessment in Software Engineering. 2014.

[S16] Greiler, Michaela, Kim Herzig, and Jacek Czerwonka. "Code ownership and software quality: A replication study." 2015 IEEE/ACM 12th Working Conference on Mining Software Repositories. IEEE, 2015.

[S17] Müller, Christoph, Guido Reina, and Thomas Ertl. "In-situ visualisation of fractional code ownership over time." 8th International Symposium on Visual Information Communication and Interaction. 2015.

[S18] Qiu, Yilin, et al. "An empirical study of developer quality." 2015 IEEE International Conference on Software Quality, Reliability and Security-Companion. IEEE, 2015.

[S19] Foucault, Matthieu, et al. "On the usefulness of ownership metrics in open-source software projects." Information and Software Technology 64 (2015): 102-112.

[S20] Faragó, Csaba, Péter Hegedűs, and Rudolf Ferenc. "Code ownership: Impact on maintainability." International Conference on Computational Science and Its Applications. Springer, Cham, 2015.

[S21] Thongtanunam, Patanamon, et al. "Revisiting code ownership and its relationship with software quality in the scope of modern code review." Proceedings of the 38th international conference on software engineering. 2016.

[S22] Lindsjørn, Yngve, et al. "Teamwork quality and project success in software development: A survey of agile development teams." Journal of Systems and Software 122 (2016): 274-286.

[S23] Businge, John, et al. "Code authorship and fault-proneness of open-source android applications: An empirical study." Proceedings of the 13th International Conference on Predictive Models and Data Analytics in Software Engineering. 2017.

[S24] Zhu, Wenhan, and Michael W. Godfrey. "Mea culpa: How developers fix their own simple bugs differently from other developers." 2021 IEEE/ACM 18th International Conference on Mining Software Repositories (MSR).

[S25] Herzig, Kim, and Nachiappan Nagappan. "The impact of test ownership and team structure on the reliability and effectiveness of quality test runs." Proceedings of the 8th ACM/IEEE International Symposium on Empirical Software Engineering and Measurement. 2014.

**Appendix (II) - Studies and Variables Explanation**

Table 15: Specifications of the papers directly relevant to "Ownership" of software artifacts included in our systematic literature review. Paper which did not discuss ownership directly (such as S66) are excluded from this table.

| Paper | What? | Who? | How? | Dependent variable | Independent variable | Control variable |
|---|---|---|---|---|---|---|
| S1 | Code, Component | Developer | Collective shared | Number of coding errors | Expertise location, Expertise aware | Developer experience, team size, and project size |
| S2 | Product | Product manager | Collective shared | N/A | N/A | N/A |
| S3 | Code (binary files) | Developer | Weighted shared | Number of post release bugs | Code Churn, Code Complexity, Code dependencies, Code coverage, Number of pre release bugs, Number of developers, Number of ex-developers, Edit frequency, Depth of master ownership, % of Org. contributing to development, Level of org. code ownership, Overall org. Ownership, Org. intersection factor | N/A |
| S4 | Code, File | Developer | Weighted shared | N/A | Code churn, Number of commits | N/A |
| S5 | Code, File | developer | Collective shared | N/A | N/A | N/A |
| S6 | Code, Component | Developer | Weighted shared | Number of post release failures | Code churn,Code complexity,Test coverage,Binary dependencies, Number of developers | Developer geographical location |
| S7 | Code, Component | Developer, Manager | Weighted shared | Number of pre release failures, Number of post release failures | Highest % of contribution a developer made to a software component, Number of major contributor, Number of minor contributor, Number of developers | Code size, Code churn, Code complexity |
| S8 | Code, file | Developer | Weighted shared | Number of line of code to fix defect | Developer experience, Collective ownership | Project size, Number of bugs linked with bug fixing revision, Number of developers |
| S9 | Issue | Developer | Weighted shared | Number of active issues | Number of developers | N/A |
| S10 | Code | Developer | Weighted shared | Collective ownership of a class | Linguistic Topics using LDA | N/A |
| S11 | Component | Organization | Weighted shared | Number of pre release failures, Number of post release failures | Number of commits | Code size, Code churn, Code complexity |
| S12 | Code (authorship) | Developer | Weighted shared | N/A | N/A | N/A |
| S13 | Code (authorship) | Developer | Dedicated | Number of bugs | Characters churn per line per commit, | N/A |
| S14 | Task,Review | Developer | Weighted shared | The time between review creation and last updating | The level of involvement of a review's owner in the overall review process | Number of patches associated with a review,The median of the approval scores received by a reviewer, The degree centrality of review in a network based on review similarity, Number of reviewers, Number of comments on a review by developers who do not own it, Number of reviews owned by its owner for each review |
| S15 | Code,Component | Developer | Weighted shared | Number of bugs | Highest % of contribution a developer made to a software component,Number of major contributor,Number of minor contributor,Number of developers | Code size, Code complexity |
| S16 | Code, Component | Developer, Manager, Organization | Collective shared | Number of bugs | Highest % of contribution a developer made to a software component, Number of developers, Number of minor contributor, Number of minimal contributor, Number of manager, Average ownership value of files in a directory, % of commits of highest contributor in a directory, % of commits of lowest contributor in a directory, Average contributors in a directory,% of minor contributor in a directory, % of minimal contributor in a directory, % of major contributor in a directory, Average of minimal contributor in a directory, Average of minor contributor in a directory, Lowest ownership value of file , Number of files with 50% ownership value | N/A |
| S17 | Code | Developer | Weighted shared | Fractional ownership | Number of change of every character in a source code file | N/A |
| S18 | Code,commit | Developer | Weighted shared | Number of bugs | Number of developers,Number of commits | N/A |
| S19 | Code, Component | Developer | Weighted shared | Number of post release bugs | Code size, Code churn, Number of developers, Number of files touched by a developer, Highest % of contribution a developer made to a software component, Number of major contributor, Number of minor contributor | N/A |
| S20 | Code | Developer | Weighted shared | N/A | N/A | N/A |
| S21 | Code, component | Developer, Reviewer | Weighted shared | Number of post release defects | Highest code ownership value, Highest review specific ownership value, Proportion of minor developer and major reviewer,Proportion of major developer and major reviewer,Proportion of major developer and minor reviewer,Proportion of minor developer and minor reviewer,Proportion of minor developer and major reviewer | Code size, Code churn, Entropy,Number of developer,Number of reviewer |
| S22 | Code | Developer | Collective | N/A | N/A | N/A |
| S23 | Code, file | Developer | Weighted Shared | Number of faults | Number of major contributor,Number of minor contributor,Number of developers, Highest% of contribution a developer made to a software component | Project size, Code complexity |
| S24 | Bug fix | Developer | Dedicated | Code churn, Bug fixing time | Authorship of bug fixing code | N/A |
| S25 | Test cases | Developer (Test Engineer), organization | Weighted shared | Number of fixed bugs, Relative number of fixed bugs per test suite execution | Number of engineers and ex-engineers who contributed to a test suite, Length of organizational communication path, Number of mailing list and user account as test owner in a test suite | N/A |
| S26 | Project | Developer, Department | Weighted shared | Number of failures per hour | Number of developers, Number of new developers, Expansion acceleration, Department modularity | N/A |
| S27 | Code, commit | Developer | Dedicated | N/A | N/A | N/A |
| S28 | Build | Developer | Dedicated, Collective shared | Build change type, Build ownership style | % of all build commits of a project that belongs to a specific category (Adaptive, Corrective, Perfective, Preventive, new functionality, Reflective) of build changes, Average amount of build churn per file changed by the commits of a build change category (Adaptive, Corrective, Perfective, Preventive, new functionality, Reflective), The median number of unique build files modified by the build commits in a build change category (Adaptive, Corrective, Perfective, Preventive, new functionality, Reflective) | N/A |
| S29 | Code,Commit | Developer | Weighted shared | Number of post-release defects | Number of developers, Number of major contributor, Number of minor contributor, Highest % of contribution a developer made to a software component | Code Size, Code complexity |
| S30 | Code, File | Developer | Weighted shared | Number of refactored files | Mean ownership computed on file,Mean ownership computed on author | N/A |
| S31 | Code | Developer | Shared | N/A | N/A | N/A |
| S32 | Product | Developer | Collective shared | Turnover Intention | Value Fit, Demands-Abilities Fit , Psychological ownership | Financial Compensation |
| S33 | Code | Developer | Collective Shared | N/A | N/A | N/A |
| S34 | Code, component | Developer | Weighted shared | N/A | N/A | N/A |
| S35 | Code, component (fragment) | Developer | Collective Shared | N/A | N/A | N/A |
| S36 | Code (authorship) | Developer | Weighted shared | N/A | Programming style and code smell metrics based on variables and natural language | N/A |
| S37 | Code (authorship) | Developer | Weighted shared | N/A | N/A | N/A |
| S38 | Requirement | Developer, User, Stakeholder | Shared | N/A | N/A | N/A |
| S39 | Code (authorship) | Developer | Dedicated | N/A | N/A | N/A |
| S40 | Code | Organization | Weighted shared | N/A | N/A | N/A |
| S41 | Code,fragment (authorship) | Developer | Shared | N/A | N/A | N/A |
| S42 | Code | Developer | Dedicated | N/A | N/A | N/A |
| S43 | Code, File, commit | Developer | Collective shared | N/A | Code size, Number of line of code of each author | N/A |
| S44 | Code (authorship) | Developer | Weighted shared | N/A | N/a | N/A |
| S45 | Code, Functions | Developer | Dedicated | N/A | N/A | N/A |
| S46 | Project | Developer,Team | Collective shared | N/A | N/A | N/A |
| S47 | Task | Developer | Weighted shared | N/A | N/A | N/A |
| S48 | Product | Developer | Dedicated | | | |
| S49 | Code | Developer | Dedicated | N/A | N/A | N/A |
| S50 | Issue | Developer | Shared | N/A | N/A | N/A |
| S51 | Code,File, Commit | Developer | Dedicated | N/A | N/A | N/A |
| S52 | Task | Developer | Weighted shared | N/A | N/A | N/A |
| S53 | Code, file | Developer | Dedicated | N/A | N/A | N/A |
| S54 | Code | Developer | Dedicated | Number of defects | Developer experience, Code size | N/A |
| S55 | File | Developer | Dedicated | N/A | N/A | N/A |
| S56 | Project | User | Shared | N/A | N/A | N/A |
| S57 | Code (source and binary) | Developer | Dedicated | N/A | N/A | N/A |
| S58 | Product | Manager | Dedicated | N/A | N/A | N/A |
| S59 | Product | Developer | Dedicated | N/A | N/A | N/A |
| S60 | Code | Developer | Shared | N/A | N/A | N/A |
| S61 | Code and Product | Developer | Collective Shared | N/A | N/A | N/A |
| S62 | Code, file | Developer | Shared | Degree of authorship of a specific developer | First authorship of developer through file creation, Number of changes made by file creator developer, Number of changes made by other developer except file creator | Project size |
| S63 | Code | Developer | Shared | N/A | N/A | N/A |
| S64 | Task | Developer | Dedicated | N/A | N/A | N/A |
| S65 | Code | Developer | Shared | N/A | Code size | N/A |
| S67 | Task, Project | Developer, Manager | Dedicated | N/A | N/A | N/A |
| S68 | Code | Developer | Weighted shared | N/A | N/A | N/A |
| S69 | Product | Manager | Dedicated | N/A | N/A | N/A |
| S70 | Code (authorship) | Developer | Weighted shared | N/A | N/A | N/A |
| S71 | Code | Developer | Shared | N/A | N/A | N/A |
| S72 | Code, File | Developer | Weighted shared | N/A | N/A | Code complexity, Developer experience, Type of project |
| S73 | Issue | Developer | Dedicated | N/A | N/A | N/A |
| S74 | Code | Developer | Collective shared | N/A | N/A | N/A |
| S76 | Code | Developer | Collective shared | N/A | N/A | N/A |
| S77 | Code | Developer | Dedicated, Collective shared | Proportion of contribution | Number of commits, Code churn, Code complexity, Ticket complexity, Pull Request complexity, Number of pull requests, Number of Tickets | N/A |
| S78 | Product | Product owner | Collective shared | N/A | N/A | N/A |
| S79 | Code | Developer | Collective shared | N/A | N/A | N/A |